\documentclass[twocolumn]{aastex63}
\usepackage{amsmath}
\usepackage{here}

\newcommand{\ring}{\mathrm{ring}}
\newcommand{\gas}{\mathrm{g}}
\newcommand{\dst}{\mathrm{d}}
\newcommand{\drift}{\mathrm{drift}}
\newcommand{\sigmag}{\Sigma_{\gas}}
\newcommand{\sigmagup}{\Sigma_{\gas,0}}
\newcommand{\sigmad}{\Sigma_{\dst}}
\newcommand{\sigmadup}{\Sigma_{\dst,0}}
\newcommand{\tstop}{t_{\mathrm{stop}}}

\newcommand{\taus}{\tau_{\mathrm{s}}}
\newcommand{\tausup}{\tau_{\mathrm{s},0}}
\newcommand{\cs}{c_{\mathrm{s}}}
\newcommand{\cd}{c_{\mathrm{d}}}
\newcommand{\hd}{H_{\mathrm{d}}}
\newcommand{\vk}{v_{\mathrm{K}}}

\newcommand{\gdl}{\mathrm{gdl}}
\newcommand{\vpp}{v_{\mathrm{pp}}}
\newcommand{\lmfp}{l_{\mathrm{mfp}}}
\newcommand{\mpar}{m_{\mathrm{p}}}
\newcommand{\rhoint}{\rho_{\mathrm{int}}}

\newcommand{\ddx}[1]{\frac{\partial #1}{\partial x}}
\newcommand{\ddr}[1]{\frac{\partial #1}{\partial r}}
\newcommand{\ddt}[1]{\frac{\partial #1}{\partial t}}

\newcommand{\sigmagmmsn}{\Sigma_{\mathrm{g,MMSN}}}

\def\msun{M_{\odot}}

\received{--}
\accepted{--}
\submitjournal{ApJ}

\shorttitle{Numerical study of coagulation instability}
\shortauthors{Tominaga et al.}

\begin{document}

\title{Nonlinear Outcome of Coagulation Instability in Protoplanetary Disks I: \\ First Numerical Study of Accelerated Dust Growth and Dust Concentration at Outer Radii}

\correspondingauthor{Ryosuke T. Tominaga}
\email{ryosuke.tominaga@riken.jp}
\author[0000-0002-8596-3505]{Ryosuke T. Tominaga}
\affiliation{RIKEN Cluster for Pioneering Research, 2-1 Hirosawa, Wako, Saitama 351-0198, Japan}

\author[0000-0001-8808-2132]{Hiroshi Kobayashi}
\affiliation{Department of Physics, Nagoya University, Nagoya, Aichi 464-8692, Japan}

\author[0000-0003-4366-6518]{Shu-ichiro Inutsuka}
\affiliation{Department of Physics, Nagoya University, Nagoya, Aichi 464-8692, Japan}
%
%
%



\begin{abstract}
Our previous linear analysis presents a new instability driven by dust coagulation in protoplanetary disks. The coagulation instability has the potential to concentrate dust grains into rings and assist dust coagulation and planetesimal formation. In this series of papers, we perform numerical simulations and investigate nonlinear outcome of coagulation instability. In this paper (Paper I), we first conduct local simulations to demonstrate the existence of coagulation instability. Linear growth observed in the simulations is in good agreement with the previous linear analysis. We next conduct radially global simulations to demonstrate that coagulation instability develops during the inside-out disk evolution due to dust growth. To isolate the various effects on dust concentration and growth, we neglect effects of backreaction to a gas disk and dust fragmentation in Paper I. This simplified simulation shows that either of backreaction or fragmentation is not prerequisite for local dust concentration via the instability. In most runs with weak turbulence, dust concentration via coagulation instability overcomes dust depletion due to radial drift, leading to the formation of multiple dust rings. The nonlinear development of coagulation instability also accelerates dust growth, and the dimensionless stopping time $\taus$ reaches unity even at outer radii ($>10\;\mathrm{au}$). Therefore, coagulation instability is one promising process to retain dust grains and to accelerate dust growth beyond the drift barrier.
\end{abstract}



\section{Introduction}\label{sec:intro}
Growth from (sub-)micron-sized dust grains to km-sized planetesimals is one long-standing problem in planet formation theory. Recently, exploration by the New Horizons spacecraft provided data that constrains planetesimal forming processes at outer radii in early solar nebula \citep{Stern2019,McKinnon2020}. The bi-lobe shape of one Kuiper Belt object, Arrokoth, indicates that the object might form via gravitational collapse of a dust clump in the presence of a gas disk. This scenario seems consistent with one proposed scenario of planetesimal formation \citep[see also][]{Nesvorny2019}: streaming instability and subsequent gravitational instability (GI) \citep[e.g., ][]{YoudinGoodman2005,Youdin2007,Johansen2007,Johansen2007nature,Bai2010a,Simon2016,Abod2019,Carrera2021}. However, recent studies show that the streaming instability is substantially stabilized by diffusion in gas turbulence \citep[][]{Chen2020,Umurhan2020}. \citet{McNally2021} investigated growth rates of the streaming instability in the presence of the turbulent diffusion with dust size distribution. They showed that even weak turbulence with $\alpha\sim10^{-5}$ can quench the streaming instability (see Figure 8 therein), where $\alpha$ is the turbulence strength \citep{Shakura1973}. Therefore, it is important to investigate other processes potentially leading to outer planetesimal formation, including secular GI \cite[e.g.,][]{Ward2000,Youdin2005a,Youdin2005b,Youdin2011,Shariff2011,Michikoshi2012,Takahashi2014,Tominaga2018,Tominaga2019,Tominaga2020,Pierens2021}.

Our previous work found a new instability driven by dust coagulation in a protoplantary disk \citep{Tominaga2021}. The instability that we named ``coagulation instability" leads to clumping of drifting dust even in the presence of the turbulent diffusion with $\alpha\sim 10^{-4}$. Another remarkable difference from the previous dust-gas instabilities is that coagulation instability can develop at tens of orbital periods even when dust-to-gas surface density ratio is so small ($\sim10^{-3}$) that the previous instabilities hardly develop.

According to the previous studies on coagulation, dust grains are depleted as a result of the inside-out coagulation and radial drift, leading to a low dust-to-gas ratio of $\sim10^{-3}$ \citep[e.g.,][]{Brauer2008,Birnstiel2009,Okuzumi2012}. This indicates that coagulation instability will operate earlier than the other instabilities proposed in the previous studies such as streaming instability or secular GI. \citet{Tominaga2021} indicates that coagulation instability accelerates dust growth and may form planetesimals directly. At the same time, dust concentration and growth via coagulation instability can also set up preferable regions for the other instabilities to subsequently develop toward planetesimal formation. Since the instability potentially has a significant contribution to planetesimal formation, the viability of coagulation instability and the above hypothesis should be tested by numerical simulations.

In this series of papers, we perform numerical simulations and investigate the nonlinear outcome of coagulation instability. This paper (Paper I) presents the first results of numerical simulations. We conduct local simulations to prove the existence of coagulation instability as predicted in our linear analysis \citep[][]{Tominaga2021}. We also conduct radially global simulations and show that coagulation instability does develop during the inside-out disk evolution via dust growth. The instability is found to cause radial dust concentration and to accelerate dust growth. In Paper I, we intentionally neglect frictional backreaction (deceleration of the drift) and dust fragmentation to isolate the contributions to dust concentration and growth from various effects. This simplification allows clear distinction between linear/nonlinear coagulation instability and self-induced dust trap process due to a combination of coagulation and backreaction \citep{Gonzalez2017}. In other words, such simplified simulations demonstrate that either of backreaction or fragmentation is not prerequisite for dust concentration via the instability. Therefore, based on this simplification, we aim to obtain clear physical understanding of nonlinear outcome of coagulation instability in Paper I. Paper II (Tominaga et al. 2022b, submitted) presents simulations of coagulation instability under the influence of backreaction and fragmentation.

Paper I is organized as follows. We describe basic equations and numerical methods in Section \ref{sec:method}. In Section \ref{sec:local}, we first briefly review the linear analysis of coagulation instability (Section \ref{subsec:review}) and show results of local simulations that we compare with the linear analysis. We present radially global simulations of coagulation instability in Section \ref{sec:global} and discuss to what extent coagulation instability concentrates dust grains before dust falls onto a central star. In Section \ref{sec:discussion}, we discuss further evolution of resulting dust rings and possible way to retain dust grains in a disk. We summarize in Section \ref{sec:summary}.

\section{Methods}\label{sec:method}
\subsection{Basic equations}\label{subsec:basiceq}
In this work, we consider dust evolution in a steady axisymmetric gas disk around a star of $1\msun$. Equation of continuity for dust in cylindrical coordinates $(r,\phi,z)$ is given by
\begin{equation}
\ddt{\sigmad}+\frac{1}{r}\ddr{ }\left(r\sigmad v_r\right)=0,\label{eq:dusteoc}
\end{equation}
\begin{equation}
v_r\equiv\left<v_r\right>-\frac{D}{\sigmad}\ddr{\sigmad},\label{eq:vr}
\end{equation}
where $\sigmad$ is the dust surface density and $D$ is the diffusion coefficient. The first term in Equation (\ref{eq:vr}) denotes a mean drift velocity of dust grains. Assuming that the gas rotates at sub-Keplerian velocity $(1-\eta)\vk$ where $\vk=\sqrt{G\msun/r}$ is the Keplerian velocity and $G$ is the gravitational constant, one obtains the radial terminal velocity in the test particle limit as follows \citep[e.g.,][]{Adachi1976,Weidenschilling1977,Nakagawa1986}:
\begin{equation}
\left<v_r\right>=-\frac{2\taus}{1+\taus^2}\eta \vk,\label{eq:meanvr_woBR}
\end{equation}
where $\taus\equiv\tstop\Omega$ is the stopping time of dust grains, $\tstop$, normalized by the Keplerian angular velocity $\Omega\equiv\vk/r=\sqrt{G\msun/r^3}$. The value of $\eta$ for a given gas density $\rho_{\gas}$ and temperature $T$ is determined by
\begin{equation}
\eta\equiv-\frac{1}{2\rho_{\gas}r\Omega^2}\frac{\partial\cs^2\rho_{\gas}}{\partial r},\label{eq:eta}
\end{equation}
where $\cs=\sqrt{k_{\mathrm{B}}T/\mu m_{\mathrm{H}}}$ is the sound speed, and $k_{\mathrm{B}}$ and $m_{\mathrm{H}}$ are the Boltzmann constant and the hydrogen mass, respectively. We adopt the mean molecular weight $\mu$ of 2.34. We use the analytical formula of the radial diffusion coefficient derived by \citet{YL2007}:
\begin{equation}
D=\frac{1+\taus+4\taus^2}{(1+\taus^2)^2}\alpha\cs H.\label{eq:diffcoef}
\end{equation}

We calculate the radial dust diffusion using the gradient $\partial\sigmad/\partial r$ \citep[e.g.,][]{Cuzzi1993} instead of $\sigmag\partial\left(\sigmad/\sigmag\right)/\partial r$ where $\sigmag$ is the gas surface density \citep[e.g.,][]{Dubrulle1995}. We note that coagulation instability is hardly affected by the choice of the gradients \citep[see Section 4.1 of][]{Tominaga2021}. The difference in the diffusion flux, 
\begin{equation}
\frac{D}{\sigmad}\frac{\partial\sigmad}{\partial r}-\frac{D\sigmag}{\sigmad}\ddr{ }\left(\frac{\sigmad}{\sigmag}\right)=\frac{D}{\sigmag}\ddr{\sigmag},
\end{equation}
is small compared to the mean drift velocity in weakly turbulent disks considered in this work. For $D\sim\alpha\cs^2/\Omega$ and $\eta\sim(\cs/\vk)^2$, one obtains
\begin{equation}
\left|\frac{D}{\sigmag}\ddr{\sigmag}\times\left<v_r\right>^{-1}\right|\sim \frac{\alpha}{\taus}.
\end{equation}
The right hand side is much smaller than unity in weakly turbulent gas disks ($\alpha\sim10^{-4}$) because the drifting dust has $\taus\sim0.1-1$. As noted by \citet{Tominaga2021}, coagulation instability creates $\sigmad$ perturbations at a radial scale of $\sim H\equiv\cs/\Omega$, where $H$ is the gas scale height. In such a case, $\sigmag^{-1}\partial\sigmag/\partial r$ is smaller than $\sigmad^{-1}\partial\sigmad/\partial r$ by a factor of $H/r$. Therefore, the difference in the diffusion flux is minor \citep[see also][]{Laibe2020}.

We describe dust size evolution via coagulation adopting the moment method \citep[][]{Estrada2008,Ormel2008,Sato2016}. The evolutionary equation of mass-dominating dust sizes is derived in \citet{Sato2016}:
\begin{equation}
\ddt{\mpar}+v_r\ddr{\mpar}=\frac{2\sqrt{\pi}a^2\Delta\vpp}{\hd}\sigmad,\label{eq:dmpdt}
\end{equation}
where $\mpar$ and $a$ are the mass and the radius of the representative dust, $\Delta\vpp$ is the collision velocity, and $\hd$ is the dust scale height. Equation (\ref{eq:dmpdt}) can be derived from Smoluchowski equation for column number density for dust, that is, a vertically integrated equation. The dust scale height is a function of $\taus$ and $\alpha$ \citep{YL2007}:
\begin{equation}
\hd(\taus)=H\left(1+\frac{\taus}{\alpha}\left(\frac{1+2\taus}{1+\taus}\right)\right)^{-1/2}.
\end{equation}
 We use turbulence-driven collision velocity $\Delta \vpp=\Delta v_{\mathrm{t}}$ given by \citet{Ormel2007} in Section \ref{sec:local} for comparison with the previous linear analyses. Especially, we assume $\Delta\vpp=\sqrt{C\alpha\taus}\cs$ with a numerical constant $C$ in Section \ref{subsec:review} for simplicity, which is valid for intermediate dust sizes \citep[see Section 3.4.2 in][]{Ormel2007}. In radially global simulations in Section \ref{sec:global}, we include Brownian motion and differential drift and settling velocities in $\Delta\vpp$:
\begin{equation} 
\Delta \vpp=\sqrt{\left(\Delta v_{\mathrm{t}}\right)^2+\left(\Delta v_{\mathrm{B}}\right)^2+\left(\Delta v_{r}\right)^2+\left(\Delta v_{\phi}\right)^2+\left(\Delta v_{z}\right)^2},
\end{equation}
where $\Delta v_{\mathrm{B}}$ is the collision velocity due to Brownian motion, $\Delta v_{r}$, $\Delta v_{\phi}$, and $\Delta v_z$ are the relative velocities between dust grains of $\taus=\tau_{\mathrm{s},1}$ and $\tau_{\mathrm{s},2}$. In the test particle limit, the azimuthal and  vertical drift velocities of one dust particle are as follows \citep[e.g.,][]{Nakagawa1986}:
\begin{equation}
v_{\phi}(\taus)=\vk-\frac{1}{1+\taus^2}\eta \vk,
\end{equation}
\begin{equation}
v_z(\taus)=-\frac{\taus}{1+\taus}z\Omega.
\end{equation}
Because our model is based on the vertically averaged equations, we adopt $z=\sqrt{2/\pi}H_{\dst,1,2}$ to calculate the vertically averaged $\Delta v_z$, where $H_{\dst,1,2}\equiv(H_{\dst}(\tau_{\mathrm{s},1})^{-2}+H_{\dst}(\tau_{\mathrm{s},2})^{-2})^{-1/2}$ \citep[see][]{Taki2021}.

We focus on only evolution of $\sigmad$ and $\mpar$, and ignore the higher moments. This simplification is valid until the onset of the runaway growth because the evolution of size distribution is well described by the evolution of the peak mass $\mpar$ \citep{Kobayashi2016}. \citet{Sato2016} formulated a closure by comparing full-size simulations and single-moment simulations. They found that using $\tau_{\mathrm{s},1}=0.5\tau_{\mathrm{s},2}$ for the turbulent and differential collision velocities well reproduces the $\mpar$ evolution from the full-size simulations. We thus adopt their formalism \citep[for the details, see][]{Sato2016}.

We use the Epstein and Stokes laws to calculate $\tstop$:
\begin{equation}
\tstop= 
\begin{cases}
\displaystyle\sqrt{\frac{\pi}{8}}\frac{\rhoint a}{\rho_{\gas}\cs}, &\displaystyle \left(\frac{a}{\lmfp}\leq\frac{9}{4}\right)\\
\displaystyle\sqrt{\frac{\pi}{8}}\frac{4\rhoint a^2}{9\rho_{\gas}\cs\lmfp}, &\displaystyle\left(\frac{a}{\lmfp}>\frac{9}{4}\right)\\
\end{cases}\label{eq:def_taus}
\end{equation}
where $\rhoint=1.4\;\mathrm{g}\:\mathrm{cm}^{-3}$ is an internal mass density of dust grains (cf. Appendix \ref{app:porosity}), $\lmfp$ is the mean free path of gas. In this work, we use the midplane gas density to calculate $\tstop$ since most of the dust grains reside around the midplane ($\hd<H$) when the radial drift motion becomes significant. Assuming the Gaussian gas density profile in the vertical direction and $\rho_{\gas}=\sigmag/\sqrt{2\pi}H$, we obtain
\begin{equation}
\taus=\frac{\pi}{2}\frac{\rhoint a}{\sigmag}\mathrm{max}\left(1,\frac{4a}{9\lmfp}\right),
\end{equation}
\citep[see also][]{Sato2016}. The Epstein law is applicable in the present work because we focus on $r\gtrsim5\;\mathrm{au}$ where the gas density is low and $\lmfp$ is large enough.

\subsection{Numerical methods}\label{subsec:nummethod}
We use the Lagrangian-cell method developed in \citet{Tominaga2018} for both local and global simulations in the following sections. The advantage to use the Lagrangian scheme is the fact that the scheme is free from numerical diffusion associated with advection, which enable us to accurately describe the linear growth of coagulation instability. Using Equations (\ref{eq:vr}) and (\ref{eq:meanvr_woBR}), we update position of $i$th cell boundary, $r_{\dst,i+1/2}$ for $i=1,\;2,\;...N-1$ where $N$ denotes the number of cells. Dust densities $\Sigma_{\dst,i}$ are assigned at cell centers $r=r_{\dst,i}$ while dust sizes $a_{i+1/2}$ are assigned at cell boundaries. Assuming cell masses to be constant in time, we update dust densities with time-variable cell widths. In the radially global simulations (Section \ref{sec:global} and Paper II), the cell positions correspond to the dust positions in the laboratory frame. On the other hand, we regard the cell positions as the coordinates in the unperturbed-dust-rest frame in the local simulations, which is explained in detail in the next section.

We adopt the operator-splitting method for time integration \citep[e.g.,][]{Inoue2008}. We first update the cell boundaries' positions $r_{\dst,i+1/2}$ with $\left<v_r(r_{\dst,i})\right>$ and $\Sigma_{\dst,i}$ by a half time step, $\Delta t/2$. We again update $r_{\dst,i+1/2}$ and $\Sigma_{\dst,i}$ with the diffusion flow velocity $-D\Sigma_{\dst,i}\partial\Sigma_{\dst,i}/\partial r$ by a half time step \citep[see][]{Tominaga2020}. For both time stepping, we use the second-order Runge-Kutta scheme\footnote{Although the Lagrangian-cell method developed in \citet{Tominaga2018} is used with a symplectic integrator in our previous studies, we adopt the Runge-Kutta method in this work because we assume the terminal velocity for dust instead of solving equations of motion.}. We then update the dust sizes by one time step using Equation (\ref{eq:dmpdt}). Finally, we update $r_{\dst,i+1/2}$ and $\Sigma_{\dst,i}$ with the mean radial velocity and the diffusion flow velocity in reverse order of the above time stepping to ensure the second order accuracy \citep[][]{Inoue2008}. 

The time step $\Delta t$ is determined by 
\begin{equation}
\Delta t=\mathrm{min}(\Delta t_{\mathrm{d}}, \Delta t_{\mathrm{diff}}, \Delta t_{\mathrm{coag}}),
\end{equation}
where $\Delta t_{\mathrm{d}}$, $\Delta t_{\mathrm{diff}}$ and $\Delta t_{\mathrm{coag}}$ are the time steps limited by the Courant-Friedrich-Levy condition for the dust drift, dust diffusion, and coagulation. When the time step becomes too short because of the diffusion, we adopt the super-time-stepping scheme to accelerate the time integration \citep[][]{Alexiades1996,Meyer2012,Meyer2014}.

\section{Local simulations}\label{sec:local}
In this section, we first briefly review physical properties of coagulation instability from the previous linear analyses \citep{Tominaga2021}. We next show results of local simulations of dust density and size evolution with initial perturbations. The simulations show exponential growth of the input perturbations. We compare the results with the previous linear analysis \citep[][]{Tominaga2021} and demonstrate the existence of coagulation instability.

\subsection{Physical properties of coagulation instability}\label{subsec:review}
Our previous study \citep{Tominaga2021} derived a dispersion relation of coagulation instability based on one-fluid and two-fluid equations. The previous study shows that coagulation instability is essentially one-fluid mode \citep[see Section 3 of][]{Tominaga2021}. We thus review one-fluid dispersion relation below. We suggest referring to \citet{Tominaga2021} for detailed derivation and more comprehensive discussions. 

The previous analyses assume the turbulence-driven coagulation and $\Delta \vpp=\sqrt{C\alpha\taus}\cs$ with $C\simeq2.3$. \citet{Tominaga2021} reduced Equation (\ref{eq:dmpdt}) into an evolutionary equation for $\taus$ based on the Epstein law. In the local shearing sheet coordinates $(x,y)=(r-R, R(\phi-\Omega t))$ around a radius $R$ \citep[][]{Goldreich1965}, the equation is
\begin{equation}
\ddt{\taus}+v_x\ddx{\taus}=\frac{\sigmad}{\sigmag}\frac{\taus}{3t_0}+\frac{\taus}{\sigmag}\ddx{\sigmag u_x}-\frac{\taus}{\sigmag}v_x\ddx{\sigmag},\label{eq:dstdt_full}
\end{equation}
where $t_0\equiv (4/3\sqrt{C})\Omega^{-1}$ and an axisymmetric disk is assumed. The first term on the right hand side is a coagulation term that increases $\taus$ in time. The second term represents the change in $\taus$ caused by the enhanced $\sigmag$ through gas compression. This term can be neglected in the one-fluid analyses with the steady gas disk. The third term represents a process where $\taus$ decreases as dust moves into an inner gas-dense region if dust size $a$ is fixed. \citet{Tominaga2021} found that the third term reduces growth rates of the instability by a factor of a few. In this section, we assume uniform $\sigmag=\sigmagup$ for simplicity and use the following equation:
\begin{equation}
\ddt{\taus}+v_x\ddx{\taus}=\frac{\sigmad}{\sigmag}\frac{\taus}{3t_0},\label{eq:dstdt_app}
\end{equation}
Note that the neglected third term is consistently taken into account in radially global simulations in the next section.

The unperturbed dust surface density is assumed to be uniform: $\sigmad=\sigmadup$. Besides, only in the linear analysis, the dimensionless stopping time in the unperturbed state is uniform and constant in time: $\taus=\tausup$. We note that the background dust size and thus $\tausup$ should increase in time at the background coagulation rate $\sigmadup/3\sigmagup t_0$ (see Equation (\ref{eq:dstdt_app})). However, coagulation instability can develop faster at short wavelengths than the background coagulation \citep[see Section 2 of][]{Tominaga2021}. Therefore, the above assumption to derive the growth rates is applicable as the lowest-order approximation. Note that we consistently treat the background evolution in dust size, and thus $\taus$, in the local and global simulations.

Using Equations (\ref{eq:dusteoc}), (\ref{eq:vr}), and (\ref{eq:dstdt_app}), one can derive a complex growth rate $n$ as a function of wavenumber $k$:
\begin{align}
n=&-ik\left<v_{x,0}\right>+\frac{1}{2}\left(\frac{\varepsilon_0}{3t_0}-Dk^2\right)\notag\\
&+\frac{1}{2}\sqrt{\left(\frac{\varepsilon_0}{3t_0}+Dk^2\right)^2-\frac{4ik\varepsilon_0}{3t_0}\left<v_{x,0}\right>\frac{1-\tausup^2}{1+\tausup^2}},\label{eq:disp_app_diff}
\end{align}
where $\varepsilon_0=\sigmadup/\sigmagup$ and $\left<v_{x,0}\right>=-2\tausup\eta R\Omega/(1+\tausup^2)$. Equation (\ref{eq:disp_app_diff}) shows $\mathrm{Re}[n]>0$ for intermediate wavenumbers. 

The physics behind coagulation instability can be clearly understood in the absence of diffusion. Setting $D=0$ in Equation (\ref{eq:disp_app_diff}), we obtain the growth rate at sufficiently high wavenumbers as follows
\begin{equation} 
\mathrm{Re}[n]\simeq \sqrt{\frac{1}{2}\frac{\varepsilon_0}{3t_0}k|\left<v_{x,0}\right>|}, \label{eq:disp_app_nodiff}
\end{equation}
where we set $1\pm\tausup^2\simeq1$ assuming $\tausup<1$. Equation (\ref{eq:disp_app_nodiff}) shows that the growth rate is determined by a geometric mean of the coagulation rate $\varepsilon_0/3t_0$ and $k|\left<v_{x,0}\right>|$. The latter represents a rate of traffic jam induced by $\taus$-perturbations at a scale of $k^{-1}$. Therefore, coagulation instability can be regarded as a positive feedback process driven by a combination of coagulation and traffic jam \citep[see Figure 1 and Section 2 of][]{Tominaga2021}.

The dust diffusion neglected in Equation (\ref{eq:disp_app_nodiff}) damps short-wavelength perturbations. As a result, coagulation instability develops the most efficiently at intermediate wavelengths \citep[see Section 4.1 of][]{Tominaga2021}.

\subsection{Setups of local simulations}

Next we explain setups and the initial condition of local simulations with a radial domain around $r=R$. In local simulations in this section, we update the cell positions according to the following perturbed drift velocity:
\begin{equation}
\delta v_x \equiv  \delta \left<v_x\right>-\frac{D}{\sigmad}\ddr{\sigmad},\label{eq:dv_for_local_sim1}
\end{equation}
\begin{equation}
\delta \left<v_x\right>\equiv-\frac{2\taus(x,t)}{1+\taus(x,t)^2}\eta R\Omega+\frac{2\tausup(t)}{1+\tausup(t)^2}\eta R\Omega,\label{eq:dv_for_local_sim2}
\end{equation}
where we calculate the dimensionless stopping time as $\taus=\tstop\Omega(R)$. This frame is the unperturbed-dust-rest frame, $X$, and thus we can clearly observe propagation of waves relative to the dust. The radial coordinates $x$ (dust positions in the laboratory frame) and $X$ have the following relation:
\begin{equation}
X=x+\int^t_0 \frac{2\tausup(t')}{1+\tausup(t')^2}\eta R\Omega dt'.\label{eq:dust-rest-frame}
\end{equation}
Note that we let dust evolve in time according to Equation (\ref{eq:dmpdt}) and thus the background stopping time $\tausup(t)$ increases in time. For comparison with the linear analyses, we also assume $\Delta \vpp=\sqrt{C\alpha\taus}\cs$ in this subsection.

\begin{figure}[tp]
	\begin{center}
	\hspace{100pt}\raisebox{20pt}{
	\includegraphics[width=0.9\columnwidth]{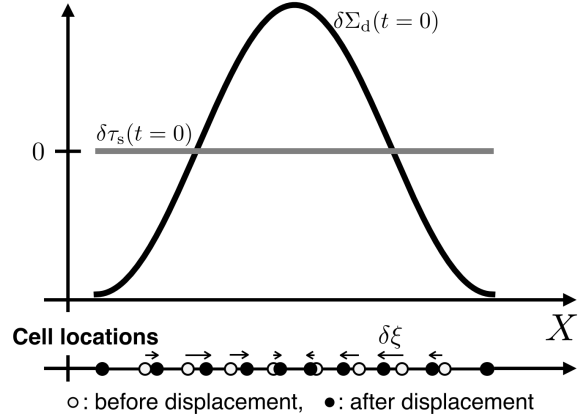}
	}
	\end{center}
	\vspace{-30pt}
\caption{Schematic summary of the initial setups of local simulations. We input sinusoidal perturbations only on dust surface density (black line), and thus $\delta\taus=0$ at $t=0$ (gray line). Since our method utilizes the Lagrangian scheme, we set up $\delta\sigmad$ displacing the cell ($\delta\xi$) as depicted on the bottom of this figure: open and filled circles represent cell locations before and after the displacement, respectively. Note that the number of cells used in the local simulations is larger than the number of filled circles in this figure.}
\label{fig:local_setup}
\end{figure}

There are four parameters: (1) $\eta R\Omega$, (2) background surface densities $\sigmagup,\;\sigmadup$, (3) initial dust size or dimensionless stopping time $\tausup$, and (4) turbulence strength $\alpha$. We adopt $\eta R\Omega\simeq54\;\mathrm{m/s}$. This choice is based on the minimum mass solar nebula (MMSN) disk model \citep[][]{Hayashi1981}. The gas surface density is uniformly set as $\sigmagup=\sigmagmmsn(20\;\mathrm{au})$ where $\sigmagmmsn(r)=53.75\;\mathrm{g\;cm}^{-2}(r/10\mathrm{au})^{-1.5}$ is the gas surface density in the MMSN disk model. We assume uniform dust-to-gas ratio of $10^{-3}$ in the local simulations. We initially adopt $\taus=\tstop\Omega(R)=10^{-2}$ throughout the domain, and set the corresponding dust sizes using Equation (\ref{eq:def_taus}). We adopt $\alpha=10^{-4}$, and $10^{-5}$ in the local simulations. The results for $\alpha=10^{-5}$ are presented in Appendix \ref{app:local_alp1e-5}.

Figure \ref{fig:local_setup} schematically shows the initial condition explained below. We simulates the time evolution of a sinusoidal perturbation. The radial width of the domain is set to be one wavelength $\lambda$ and divided into $N=128$ cells. We adopt the periodic boundary condition: the boundaries of the domain move so that $\sigmad$ becomes periodic. We stop simulations once the cell boundary of 1st cell, $r_{\dst,3/2}$, crosses the initial position of the inner domain boundary $r_{\mathrm{in}}$.

\begin{figure}[tp]
	\begin{center}
	\hspace{100pt}\raisebox{20pt}{
	\includegraphics[width=0.9\columnwidth]{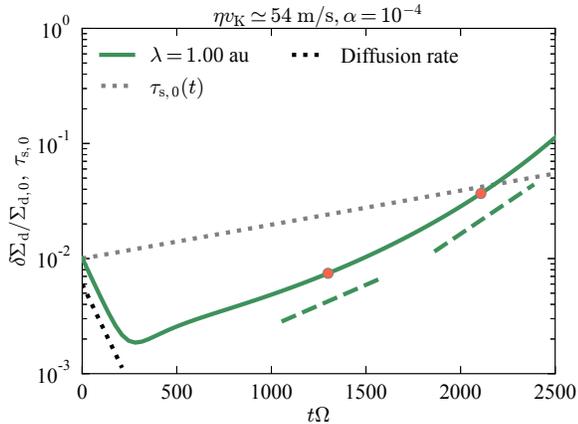}
	}
	\end{center}
	\vspace{-30pt}
\caption{Results of the local simulation with $\lambda=1\;\mathrm{au}$, $\sigmadup/\sigmagup=10^{-3}$ and $\alpha=10^{-4}$. The green solid line shows the amplitude of the dust surface density perturbation as a function of $t\Omega$. The gray dotted line shows time evolution of the background stopping time $\tausup$. The initial dust surface density perturbation is damped by the diffusion at a rate of $Dk^2$ (the black dotted line). The perturbation exponentially grows after the initial damping. The green dashed lines show the growth rates derived in the previous linear analyses (Equation (\ref{eq:disp_app_diff})). We use physical values at time marked by the orange filled circles to derive the growth rate. The linear analyses are in good agreement with the simulation at each point.}
\label{fig:local_1au_growth}
\end{figure}

Since our code is based on the Lagrangian scheme, we can set up density perturbations by displacing the cell locations (see the bottom part of Figure \ref{fig:local_setup}). We adopt the following displacement $\delta\xi$:
\begin{equation}
\delta\xi\equiv 10^{-2}\frac{\lambda}{2\pi}\sin\left(\frac{2\pi}{\lambda}(r-r_{\mathrm{in}})\right).
\end{equation}
This gives a sinusoidal perturbation on the dust surface density with one percent amplitudes. We input only dust surface density perturbations at $t=0$. The perturbations in $\sigmad$ naturally induce perturbations in $\taus$ in the subsequent time evolution.

\begin{figure*}[htp]
	\begin{center}
	\hspace{100pt}\raisebox{20pt}{
	\includegraphics[width=1.8\columnwidth]{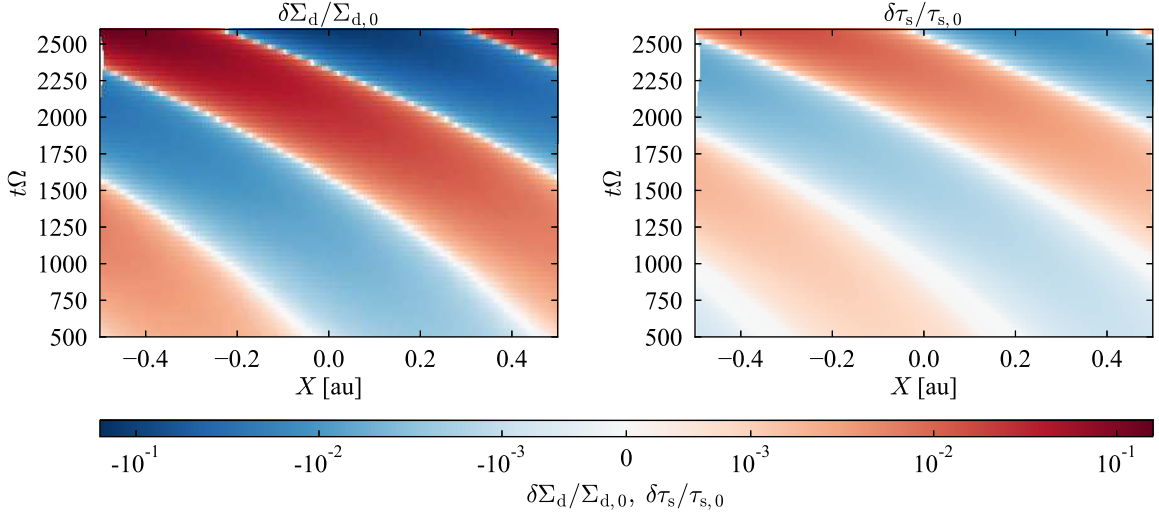}
	}
	\end{center}
	\vspace{-30pt}
\caption{Growth and propagation of $\delta\sigmad/\sigmadup$ (the left panel) and $\delta\tausup/\tausup$ (the right panel) obtained from the local simulation with $\lambda=1\;\mathrm{au}$ and $\alpha=10^{-4}$. Note that the horizontal axis is the radial position $X$ in the background-dust-rest frame (Equation (\ref{eq:dust-rest-frame})). The phase shift between $\delta\sigmad$ and $\delta\taus$ and the radial propagation of the perturbations are consistent with the previous linear analyses \citep{Tominaga2021}.}
\label{fig:dsigd_dtau_map}
\end{figure*}

\begin{figure}[htp]
	\begin{center}
	\hspace{100pt}\raisebox{20pt}{
	\includegraphics[width=0.9\columnwidth]{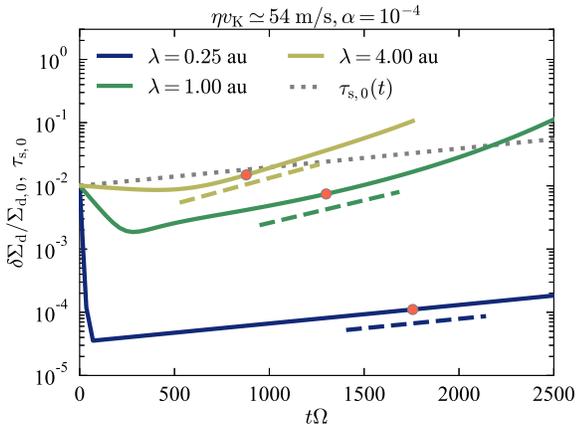}
	}
        \end{center}
	\vspace{-30pt}
\caption{Results of the local simulation with $\alpha=10^{-4}$ for $\lambda=0.25\;\mathrm{au},\;1\;\mathrm{au}$, and $4\;\mathrm{au}$. The blue, green and yellow solid lines show the evolution of $\delta\sigmad/\sigmadup$ for each wavelength. The gray dotted line shows time evolution of the background stopping time $\tausup$. The blue, green and yellow dashed lines show the growth rates derived in the previous linear analyses (Equation (\ref{eq:disp_app_diff})). We use physical values at time marked by the orange filled circles to derive the growth rate. The linear analyses have a good agreement with each simulation.}
\label{fig:local_1au_growth_lambdas}
\end{figure}

\subsection{Results and comparison with the linear analyses}

Figure \ref{fig:local_1au_growth} shows time evolution of amplitude of the dust surface density perturbation for $\lambda=1\;\mathrm{au}$ (the green solid line). The evolution of the background stopping time $\tausup(t)$ is also plotted with the gray dotted line. The dust surface density perturbation is first damped by the diffusion. After the initial damping, the perturbation exponentially grows at a larger rate than the coagulation rate (see the gray dotted line). The growth rate seen in the simulation slightly changes in time. Using Equation (\ref{eq:disp_app_diff}), we estimate linear growth rates of coagulation instability at time marked by the orange filled circles in Figure \ref{fig:local_1au_growth}. The green dashed lines show the estimated growth rates. These are in good agreement with the exponential growth rate observed in the simulation. This result thus demonstrates the existence of coagulation instability predicted by our previous analysis \citep{Tominaga2021}. Linear growth rates are determined by four parameters $\eta R\Omega$ $\varepsilon_0$, $\tausup$, and $\alpha$, and only the stopping time $\tausup$ is time-dependent. We thus attribute the observed time-dependent growth rate to the background evolution of dust sizes.

Figure \ref{fig:dsigd_dtau_map} shows time evolution of $\delta\sigmad/\sigmadup$ and $\delta\tausup/\tausup$ as a function of the radial position and time. As predicted by \citet{Tominaga2021}, the dust size perturbation is shifted outward relative to the surface density perturbation \citep[see Section 2.2 of][]{Tominaga2021}. Note that the dust cells move at the perturbed velocity $\delta v_x$, and we extracted the background drift motion $\left<v_{x,0}\right>$ (Equations (\ref{eq:dv_for_local_sim1}) and (\ref{eq:dv_for_local_sim2})). Therefore, the inward radial propagation of the perturbations in Figure \ref{fig:dsigd_dtau_map} means that the phase speed is larger than the background drift speed, which is also consistent with the results in \citet{Tominaga2021} (see also Equation (\ref{eq:disp_app_diff})).

Figure \ref{fig:local_1au_growth_lambdas} compares the results of simulations for three different wavelengths: $\lambda=0.25\;\mathrm{au},\;1\;\mathrm{au}$, and $4\;\mathrm{au}$. Shorter wavelength perturbations experience more significant initial damping due to the diffusion (e.g., the blue solid line). At later time, each simulation shows exponential growth consistent with the previous linear analyses (see dashed lines), which in turn validates our numerical scheme. 

We note that the linear analyses ignored the background evolution of the stopping time. Regardless of this assumption, the growth rates from the linear analyses well reproduce the exponential growth rates observed in the local simulations. This in turn validates the assumption in the linear analyses.

\section{Radially global simulations}\label{sec:global}
In this section, we show results of radially global simulations. We investigate (1) whether the instability develops into nonlinear growth phase before dust grains fall onto a central star, and $\taus$ locally increases up to unity, and (2) how much dust surface density and dust-to-gas ratio increase. 

\begin{deluxetable*}{c|ccccccc}[ht]
\tablecaption{Summary of parameters and results\label{tab:param}}
\tablehead{
\colhead{Runs} & \colhead{$q$} & \colhead{$\frac{\Sigma_{\gas,10}}{\Sigma_{\gas,\mathrm{MMSN}}(10\;\mathrm{au})}$} & \colhead{$T_{100}$ [K]} & \colhead{$\alpha$} & \colhead{$r_{\mathrm{in}}$ [au]} & \colhead{$\taus\to 1$ via CI?} & \colhead{$(r_{\ring},\;\Sigma_{\dst,\ring}/\sigmag$)}
}
\startdata
q3T28a1 & 3 & 1 & 28  & $1\times10^{-4}$ & 5  & Yes & $(34.9\;\mathrm{au},\;2.80\times 10^{-1})$\\
q3T20a1 & 3 & 1 & 20  & $1\times10^{-4}$ & 5  & Yes & $(37.7\;\mathrm{au},\;2.96\times 10^{-1})$\\
q3T10a1 & 3 & 1 & 10  & $1\times10^{-4}$ & 5  & Yes & $(40.1\;\mathrm{au},\;1.65\times 10^{-1})$\\
q2T28a1 & 2 & 1 & 28  & $1\times10^{-4}$ & 5  & Yes & $(31.5\;\mathrm{au},\;1.67\times 10^{-1})$\\
q2T20a1 & 2 & 1 & 20  & $1\times10^{-4}$ & 5  & Yes & $(32.4\;\mathrm{au},\;1.50\times 10^{-1}) $\\
q2T10a1 & 2 & 1 & 10  & $1\times10^{-4}$ & 5  & Yes & $(39.7\;\mathrm{au},\;1.23\times 10^{-1})$\\
q1T28a1* & 1 & 0.5 & 28  & $1\times10^{-4}$ & 20  & Yes & $(30.3\;\mathrm{au},\;1.48\times 10^{-1})$\\
q1T20a1* & 1 & 0.5 & 20  & $1\times10^{-4}$ & 20  & Yes & $(33.9\;\mathrm{au},\;1.40\times 10^{-1})$\\
q1T10a1* & 1 & 0.5 & 10  & $1\times10^{-4}$ & 20  & Yes & $(33.6\;\mathrm{au},\;5.42\times 10^{-2})$\\
q3T20a3 & 3 & 1 & 20  & $3\times10^{-4}$ & 5  & Yes & $(13.6\;\mathrm{au},\;1.00\times 10^{-1})$\\
q3T20a5 & 3 & 1 & 20  & $5\times10^{-4}$ & 5  & Yes & $(6.74\;\mathrm{au},\;7.25\times 10^{-2})$\\
q2T20a3 & 2 & 1 & 20  & $3\times10^{-4}$ & 5  & Yes & $(16.4\;\mathrm{au},\;7.50\times 10^{-2})$\\
q2T20a5 & 2 & 1 & 20  & $5\times10^{-4}$ & 5  & Yes & $(9.22\;\mathrm{au},\;6.24\times 10^{-2})$\\
q1T20a1 & 1 & 0.5 & 20  & $1\times10^{-4}$ & 5  & Yes & $(27.3\;\mathrm{au},\;9.70\times 10^{-2})$\\
q1T20a3 & 1 & 0.5 & 20  & $3\times10^{-4}$ & 5  & Yes & $(8.72\;\mathrm{au},\;1.22\times 10^{-2})$\\
q1T20a5 & 1 & 0.5 & 20  & $5\times10^{-4}$ & 5  & Yes & $(7.15\;\mathrm{au},\;8.99\times 10^{-3})$\\
q1T20a3* & 1 & 0.5 & 20  & $3\times10^{-4}$ & 20  & No & -\\
q1T20a5* & 1 & 0.5 & 20  & $5\times10^{-4}$ & 20  & No & -\\
\enddata
\tablecomments{``CI" on the top of 7th column stands for coagulation instability. The dust-to-gas ratio at the resulting rings (8th column) is derived from the last time step of each simulation. }
\end{deluxetable*}

\subsection{Disk models}
We assume the following gas surface density and temperature profile:
\begin{equation}
\sigmag(r)=\Sigma_{\gas,10}\left(\frac{r}{10\;\mathrm{au}}\right)^{-q/2},
\end{equation}
\begin{equation}
T(r) = T_{100}\left(\frac{r}{100\;\mathrm{au}}\right)^{-1/2},
\end{equation}
where $\Sigma_{\gas,10},\; q$ and $T_{100}$ are constant parameters. We assume initial dust-to-gas ratio of $10^{-2}$ in all runs:
\begin{equation}
\sigmad(r)=10^{-2}\Sigma_{\gas,10}\left(\frac{r}{10\;\mathrm{au}}\right)^{-q/2}.
\end{equation}
Initial dust size is assumed to be $10\;\mu\mathrm{m}$. The choice of the initial dust sizes insignificantly affects background evolution unless the dimensionless stopping time is much less than unity because dust grains initially just grow in size at the initial location.

\begin{figure*}[htp]
	\begin{center}
\raisebox{20pt}{
	\includegraphics[width=1.8\columnwidth]{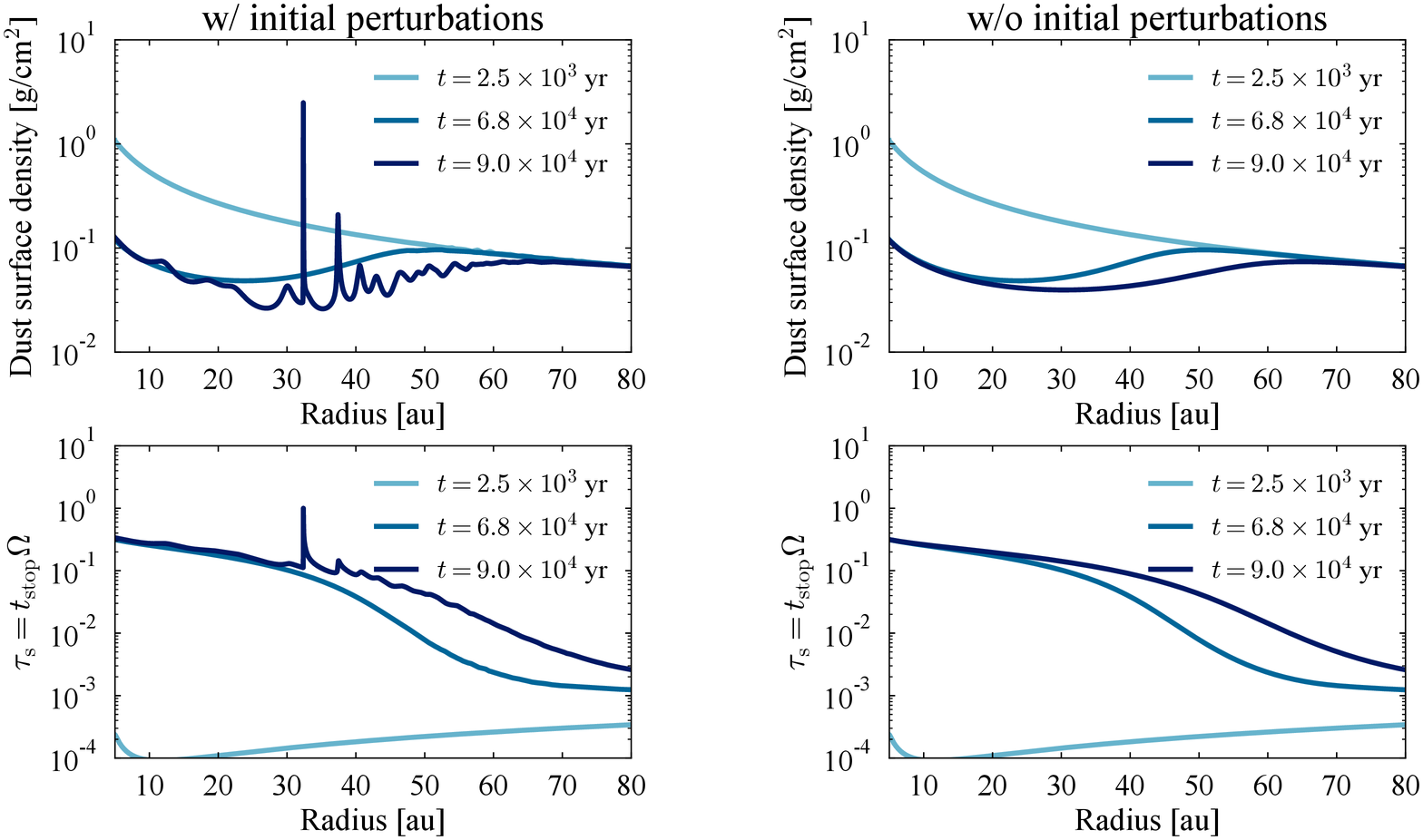}
		}
	\end{center}
	\vspace{-30pt}
\caption{Time evolution of dust surface density (top panels) and dimensionless stopping time (bottom panels) obtained from the q2T20a1 run. The left two panels and the right two panels show evolution with and without initial perturbations, respectively. Coagulation instability locally increases dust-to-gas ratio up to $\sim10^{-1}$. We stop the simulation at $t=9.0\times10^4\;\mathrm{yr}$ since the dimensionless stopping time $\taus$ becomes larger than unity around the most collapsed dust ring.}
\label{fig:q2T20a1_sigd_st}
\end{figure*}

\begin{figure}[htp]
	\begin{center}
\raisebox{20pt}{
	\includegraphics[width=0.9\columnwidth]{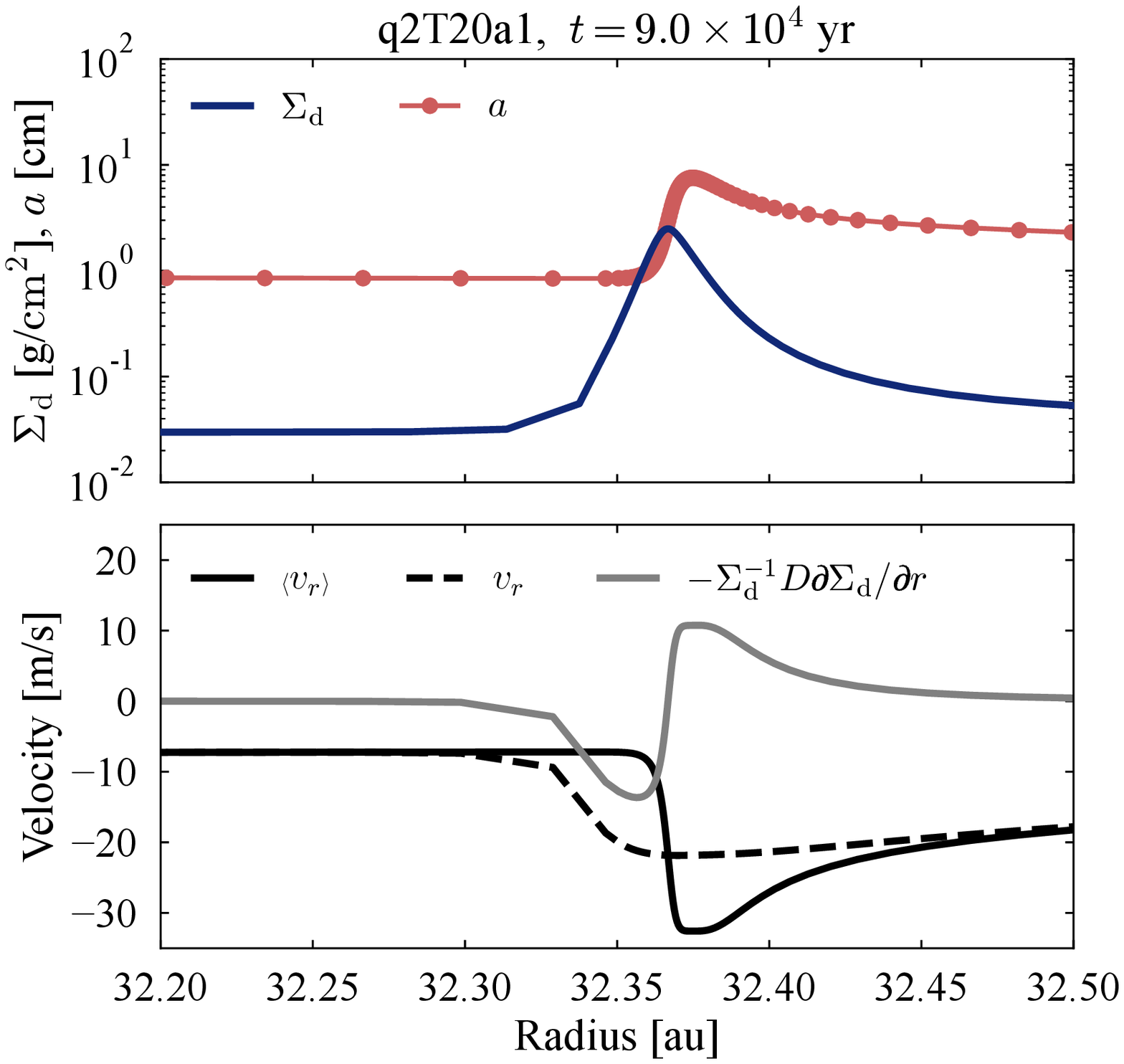}
	}	
	\end{center}
	\vspace{-30pt}
\caption{Radial profile of $\sigmad$ and dust size $a$ (top panel), and radial velocities $\left<v_r\right>$, $-\sigmad^{-1}D\partial\sigmad/\partial r$ and $v_r=\left<v_r\right>-\sigmad^{-1}D\partial\sigmad/\partial r$ (bottom panel) around the most collapsed ring at the final time step. Red filled circles show the position of the cell boundaries. One can see that the ring structure is well resolved even though the radial width is less than $0.1\;\mathrm{au}$. We find that radial gradient in $v_r$ around the left-side tail of $\sigmad$ profile ($32.3\;\mathrm{au}\lesssim r\lesssim 32.35\;\mathrm{au}$) while $v_r$ is almost uniform within the ring. Thus, the ring will sweep inner dust grains up, and the surface density increases through further evolution.}
\label{fig:q2T20a1_vr}
\end{figure}

The parameter $q$ on the exponent of the surface density profile takes three different values: $q=1,\;2,$ and $3$. In most runs, we set $\Sigma_{\gas,10}=\Sigma_{\gas,\mathrm{MMSN}}(10\;\mathrm{au})$. We adopt smaller $\Sigma_{\gas,10}$ for $q=1$ so that the gas disks are stable to nonaxisymmetric gravitational instability in the domain: $Q\equiv\cs\Omega/\pi G\sigmag>2$. 

We next set up initial perturbations. As in the local simulations (Section \ref{sec:local}), we displace the cells and input perturbations only in dust surface density at $t=0$. The displacement $\delta\xi$ adopted in the radially global simulations is the following (cf. Figure \ref{fig:local_setup}):
\begin{equation}
\delta\xi=\xi_0\sum_{j=0}^{128}\cos\left(\frac{2\pi j r}{r_{\mathrm{out}}-r_{\mathrm{in}}}+2\pi\phi_{j}\right),
\end{equation}
where $\xi_0=0.2(r_{\mathrm{out}}-r_{\mathrm{in}})/N$, $r_{\mathrm{out}}$ is the outer boundary radius, and $\phi_{j}$ is random number less than unity. As described below, we adopt the radial domain size of $95\;\mathrm{au}$ in most runs with $N=2^{14}=16384$. Thus, the 128th mode included in $\delta \xi$ has a wavelength less than $1\;\mathrm{au}$. The above displacement gives initial $\sigmad$-perturbations of about 10-20 percents at most, but dust diffusion decreases the perturbation amplitudes down to one percent or a few percents before the onset of coagulation instability. We note that the adopted $\delta\xi$ in the present simulations is small enough that cell crossing by the initial displacement does not occur.

To see wave propagation during the development of the instability, we input initial perturbations only in the outer half region reducing the displacement $\delta\xi$ in all runs as follows:
\begin{equation}
\delta\xi\to\delta\xi\times\frac{1}{2}\left[1+\tanh\left(\frac{r-R_0}{\Delta r}\right)\right],
\end{equation}
where $R_0=50\;\mathrm{au}$ and $\Delta r=1\;\mathrm{au}$. If we input perturbations without the reduction, we observe fast growth of coagulation instability at inner radii and dust growth up to $\taus=1$ well before outer dust grains start drifting because the timescales of the instability are proportional to $\Omega^{-1}\propto r^{3/2}$ (see also Appendix \ref{app:Rc}). 

Labels of runs, parameters, the adopted initial perturbation are listed in Table \ref{tab:param}.

\subsection{Numerical setups}
The number of cells $N$ is $2^{14}=16384$. The domain is uniformly spaced by $N$ cells at $t=0$. The main reason to adopt this large number of cells is that the nonlinear evolution of the instability results in very narrow structures whose width is less than 0.1 au as shown in the next subsection. Note that the Lagrangian scheme automatically increases the spatial resolutions of high density regions. However, coagulation instability develops after dust grains are depleted as a result of the radial drift, meaning that the resolution of the background state first becomes lower than the initial resolution. Therefore, to describe nonlinear coagulation instability, we need the relatively large number of cells even when we utilize the Lagrangian scheme. We note again that the Lagrangian scheme is free from numerical diffusion regardless of the number of cells, and thus this is still the advantage to use the Lagrangian scheme.

The inner and outer boundaries are set initially at $r_{\mathrm{in}}=5\;\mathrm{au}$ and $r_{\mathrm{out}}=100\;\mathrm{au}$ in most runs. As a disk evolves and dust grows, the cells move inward and flow out of the domain. We thus let the inner boundary move with the 1st cell boundary while the outer boundary radius is fixed. We discuss disk evolution in $r_{\mathrm{in}}(t=0)\leq r\leq r_{\mathrm{out}}$.

We adopt $r_{\mathrm{in}}=20\;\mathrm{au}$ at $t=0$ for a run with $T_{100}=10\;\mathrm{K}$ and $q=1$ because even background dust evolution results in dust sizes of $\taus=1$ at inner radii (q1T10a1* run). We also adopt $r_{\mathrm{in}}=20\;\mathrm{au}$ in other runs to investigate parameter dependence under the same configuration of the domain (the q1T28a1* and q1T20a1* runs) and to check the $r_{\mathrm{in}}$-dependence of the results (the q1T20a3* and q1T20a5* runs).

The moment method we utilize in this work might be valid for $\taus\leq 1$ because dust grains of $\taus=1$ is the most significantly depleted because of the fastest drift speed, and dust size distribution will become bimodal if there are larger solids of $\taus>1$. Therefore, we stop simulations once the dimensionless stopping time becomes larger than unity as a result of coagulation instability. The simulations otherwise last for $2\times10^5\;\mathrm{yr}$.

\subsection{Results}
The results are briefly summarized in 7th and 8th columns of Table \ref{tab:param}. On the 8th column, we list radial locations of dust surface density maximum, $r_{\ring}$, and resulting dust-to-gas ratios, $\Sigma_{\dst,\ring}/\sigmag$, at the time when $\taus=1$ is achieved. We find that in most runs coagulation instability develops into nonlinear growth phase before perturbations drift into the outside of the domain, locally increasing $\taus$ up to unity. We also find that dust-to-gas surface density ratio can increase at least by an order of magnitude after dust depletion down to $\sigmad/\sigmag\sim10^{-3}$. 

\begin{figure*}[htp]
	\begin{center}
	\hspace{100pt}\raisebox{20pt}{
	\includegraphics[width=1.8\columnwidth]{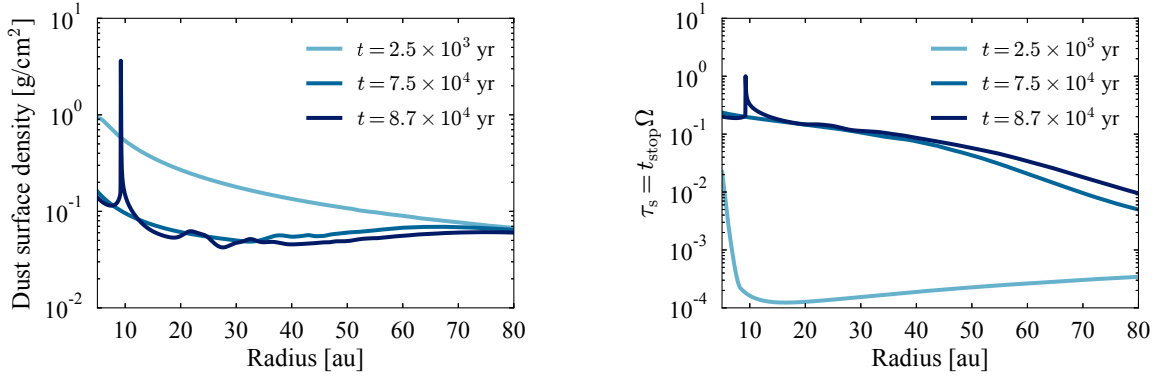}
	}
        \end{center}
	\vspace{-30pt}
\caption{Time evolution of dust surface density (the left panel) and dimensionless stopping time (the right panel) obtained from the q2T20a5 run. As in the q2T20a1 run, we observe a local increase in dust-to-gas ratio by an order of magnitude. The dimensionless stopping time locally becomes larger than unity at $t=8.7\times10^4 \;\mathrm{yr}$.}
\label{fig:q2T20a5_sigd_st}
\end{figure*}

Figure \ref{fig:q2T20a1_sigd_st} shows the results of the q2T20a1 run. The left two panels show the time evolution of dust surface density and dimensionless stopping time. We also show the evolution in the absence of the initial perturbation on the right two panels. As mentioned in the previous studies, inside-out dust coagulation and radial drift result in dust depletion and dust-to-gas ratio of order $10^{-3}$ \citep[e.g.,][]{Brauer2008}. In the presence of the initial perturbations, they exponentially grow via coagulation instability, and dust-to-gas ratio locally increases up to $\sim10^{-1}$. The dimensionless stopping time increases above unity around the most collapsed dust ring at $t=9.0\times10^4\;\mathrm{yr}$. 

Regardless of the presence of the radial dust diffusion, we find that the nonlinear coagulation instability creates narrow rings. This is because the radial concentration is driven by the radial drift velocity $\sim\taus\eta\vk$ that is much larger than $D/r$ by a factor of $\taus/\alpha$. The difference in the drift velocity can increase the dust surface density significantly until the density gradient $\partial\sigmad/\partial r$ becomes large enough that the dust diffusion balances with the drift motion.

Figure \ref{fig:q2T20a1_vr} shows dust surface density, dust sizes and radial velocity profiles around the most collapsed ring in the q2T20a1 run. Dust sizes are 1-10 cm within the ring in this run. We note that very narrow ring structures with width of $\sim 0.05\;\mathrm{au}$ are well resolved in our simulations. The bottom panel of Figure \ref{fig:q2T20a1_vr} shows that there is a radial gradient in $v_r$ in $32.30\;\mathrm{au}\lesssim r\lesssim 32.35\;\mathrm{au}$ while $v_r$ is almost uniform around the $\sigmad$-peak radius (the dashed line). Similar velocity profiles are found in other runs. This indicates that the ring will sweep inner dust grains up with moving inward, and the dust surface density in the ring will increase through further evolution until significant dust growth results in $\taus\gg1$ and quenches the dust drift. Figure \ref{fig:q2T20a1_sigd_st} shows that there are also some rings that are still growing at $r\gtrsim 37\;\mathrm{au}$. Therefore, it is expected that further evolution will lead to dust retention at multiple locations with $\taus\geq1$.

\begin{figure*}[htp]
	\begin{center}
\raisebox{20pt}{
	\includegraphics[width=1.8\columnwidth]{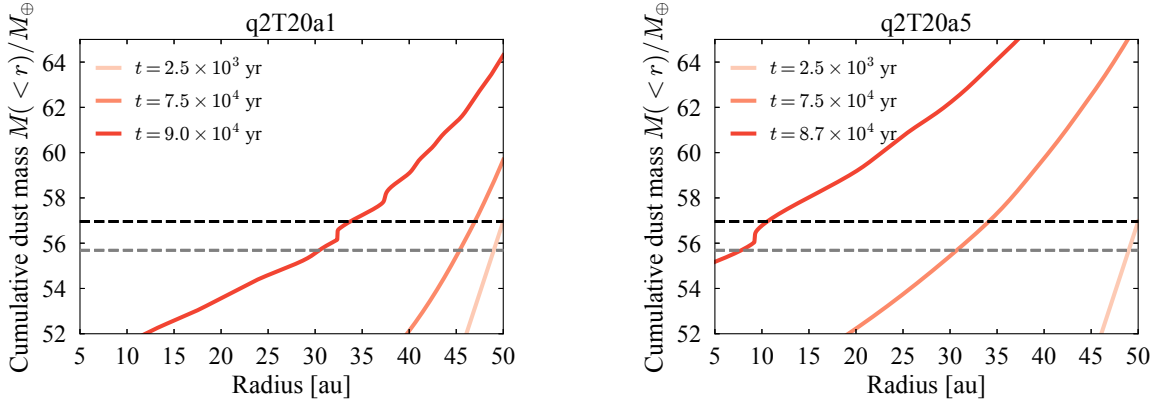}
	}	
	\end{center}
	\vspace{-30pt}
\caption{Cumulative dust mass profile from the q2T20a1 run (the left panel) and the q2T20a5 run (the right panel). The ring locations $r_{\ring}$ corresponds to ``cliffs" in the cumulative mass profiles. The gray dashed line and the black dashed line show $M(<49\;\mathrm{au})$ and $M(<50\;\mathrm{au})$ at $t=0\;\mathrm{yr}$, respectively. One can see that the resulting dust rings in both runs consist of dust cells initially located around $50\;\mathrm{au}$.}
\label{fig:cumumass_q2T20a1a5}
\end{figure*}

Figures \ref{fig:q2T20a5_sigd_st} shows results of the q2T20a5 run. Diffusion with larger $\alpha$ quenches short-wavelength modes that will exponentially grow for smaller $\alpha$. Nevertheless, coagulation instability locally increases $\taus$ up to unity as well as in the q2T20a1 run. The most collapsed ring in the q2T20a5 run is located at smaller radii than in the q2T20a1 run. The reason of this is smaller growth rates of coagulation instability in the q2T20a5 run because of stronger turbulent diffusion. Indeed, in smaller-domain-size simulations (q1T20a3* and q1T20a5*) perturbations flow out across $r=r_\mathrm{in}(t=0)$ before a significant increase in their amplitudes.  One can see that dust grains initially located at the perturbed region drift into smaller radii in Figure \ref{fig:cumumass_q2T20a1a5}. Figure \ref{fig:cumumass_q2T20a1a5} shows time evolution of cumulative mass profiles for the q2T20a1 and q2T20a5 runs. The cumulative mass also includes mass of cells flowing out of the domain. Because of the mass conservation, one can see the motion of cells from Figure \ref{fig:cumumass_q2T20a1a5}. For instance, a dust cell initially at $r=50\;\mathrm{au}$ is located at intersections of black dashed line and each cumulative mass profiles. We also plot a line which shows the cumulative mass of $M(<49\;\mathrm{au})$ at $t=0\;\mathrm{yr}$ (gray dashed line) as a reference. The positions of the dust rings in each run correspond to positions of ``cliffs" on the cumulative mass profiles. The rings in both runs consist of dust cells initially located around $r\simeq 49-50\;\mathrm{au}$. We thus attribute the difference in $r_{\ring}$ to the difference in radial distance over which dust drifts within the growth time of coagulation instability\footnote{The background evolution in the q2T20a5 run is slightly faster than in the q2T20a1 run when dust grains are small and $\alpha>\taus$. We find that difference in the dust scale height leads to the difference in the dust growth time in such a case. Note that when dust grains become large ($\alpha\gg\taus$) and we can approximate the dust scale height as $\hd\simeq\sqrt{\alpha/\taus}H$, the coagulation timescale is hardly dependent of $\alpha$ \citep[see also][]{Brauer2008,Okuzumi2012,Sato2016}. In particular, the coagulation timescale is independent from $\alpha$ for the turbulence-driven coagulation ($\Delta\vpp=\Delta v_{\mathrm{t}}\propto\sqrt{\alpha}\cs$).}. This interpretation is consistent with our previous linear analysis that shows the phase velocity is roughly given by the dust drift velocity.

\begin{figure*}[htp]
	\begin{center}
\raisebox{20pt}{
	\includegraphics[width=1.8\columnwidth]{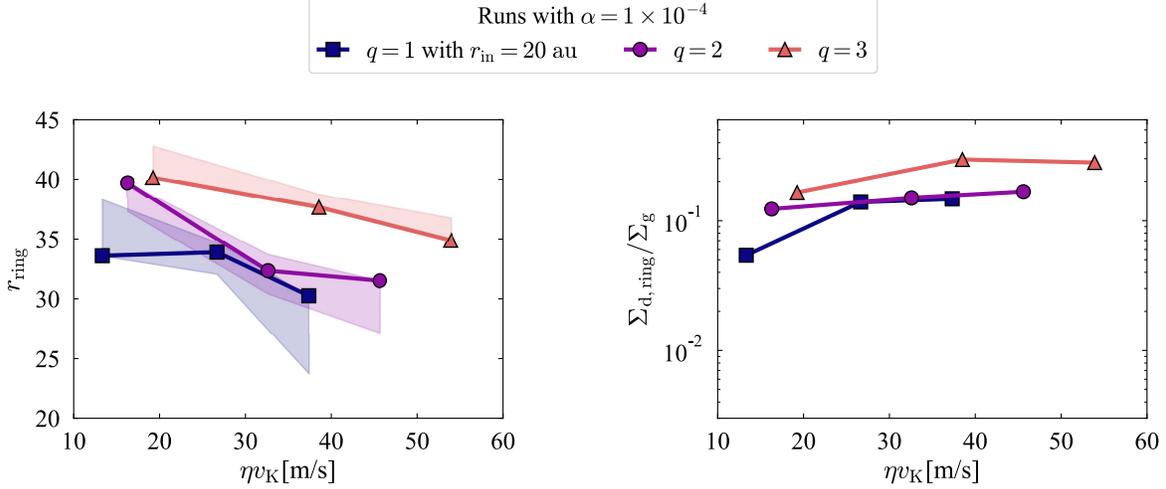}
	}
	\end{center}
	\vspace{-30pt}
\caption{Dependence of $r_{\ring}$ (left panel) and $\Sigma_{\dst,\ring}/\sigmag$ (right panel) on $\eta \vk$. On both panels, solid lines with marks show the resulting $r_{\ring}$ and $\Sigma_{\dst,\ring}/\sigmag$. On the left panel, we plot the radial location of dust cells that are initially located in $49\;\mathrm{au}\leq r\leq 50\;\mathrm{au}$ with the shaded regions. We find that larger $\eta \vk$ makes the ring radius smaller. This is because dust grains with a certain $\taus$ move faster for larger $\eta \vk$.}
\label{fig:etavk_dependence}
\end{figure*}

\begin{figure*}[htp]
	\begin{center}
	\hspace{100pt}\raisebox{20pt}{
	\includegraphics[width=1.8\columnwidth]{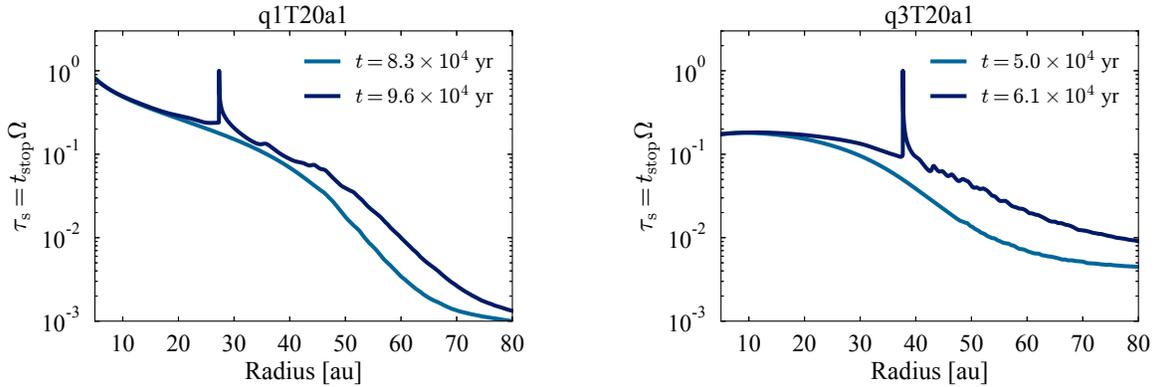}
	}
	\end{center}
	\vspace{-30pt}
\caption{Radial profiles of $\taus$ in the q1T20a1 run (the left panel) and the q3T20a1 run (the right panel). In each run, $\taus$ reaches unity in the most collapsed ring at $t=9.6\times 10^4\;\mathrm{yr}$ and $6.1\times 10^4\;\mathrm{yr}$, respectively. The background $\taus$ increases as $r$ decreases in the q1T20a1 run and becomes larger than the background $\taus$ in the q3T20a1 run. This means that dust grains drift inward more efficiently in the former run, which explains the trend that $r_{\ring}$ is smaller for smaller $q$.}
\label{fig:taus_q1T20a1_q3T20a}
\end{figure*}

We investigate $\eta\vk$-dependence of the ring position and resulting dust-to-gas ratio by changing $T_{100}$ for a given $q$ and $\alpha=1\times10^{-4}$. Figure \ref{fig:etavk_dependence} summarizes the results (solid lines with marks). On the left panel, the shaded regions show the final positions of dust cells that are initially located at $49\;\mathrm{au}\leq r\leq50\;\mathrm{au}$. All marks are in the shaded regions. This means that perturbations initially located around the inner edge of the perturbed region ($r\simeq50\;\mathrm{au}$) have enough time to develop via coagulation instability toward the nonlinear phase; the ring would consist of dust cells initially located at $r\gg50\;\mathrm{au}$ if the growth timescale was much longer than the drift timescale. We find that when $q$ is fixed the radial position $r_{\ring}$ decreases as $\eta\vk$. Since the shaded regions show the similar trend, we attribute the $\eta\vk$-dependence seen for a given $q$ to the difference in the drift velocity: dust grains with a certain $\taus$ have faster drift speed for larger $\eta\vk$ and reach inner regions. 

The left panel of Figure \ref{fig:etavk_dependence} also shows that in most cases $r_{\ring}$ is smaller for smaller $q$ although smaller-$q$ runs show lower value of $\eta\vk$ (see dark blue and orange lines). This trend originates from the radial profile of the background $\taus$. Figure \ref{fig:taus_q1T20a1_q3T20a} shows the radial profile of $\taus$ in the q1T20a1 and q3T20a1 runs. At the final time step (dark blue lines), background $\taus$ reaches the drift-limited value at inner radii. The q1T20a1 run shows $r$-dependent background $\taus$ while the q3T20a1 run shows almost uniform background $\taus$. The background $\taus$ thus becomes larger in the q1T20a1 run than in the q3T20a1 run, meaning that the drift velocity proportional to $\taus\eta\vk$ can become larger for $q=1$ even though $\eta\vk$ is smaller. Therefore, dust grains and perturbations move inward more efficiently for smaller $q$, which explains $q$-dependence in Figure \ref{fig:etavk_dependence}.

The resulting $\Sigma_{\dst,\ring}/\sigmag$ is almost similar in all run although we can see slightly larger value for larger $\eta\vk$ and larger $q$. This slight dependence is also explained based on the difference in the $\taus$ profile as in the $q$-dependence on the left panel\footnote{The difference in initial perturbations also affect to some extent.}. Larger $\eta\vk$ reduces the background drift-limited $\taus$ because of the higher drift speed. In such a case, coagulation instability locally increases dust surface density more before $\taus$ in rings reaches unity via the instability and we stop the simulations. If $q$ is small, the background $\taus$ becomes larger especially at inner radii as mentioned above, and there is little space for $\taus$ perturbations to develop up to unity, leading to smaller dust surface density.

\section{Discussions}\label{sec:discussion}
\subsection{On possible ways to retain dust grains in a disk}
As shown above, we observe significant dust concentration into a ring via coagulation instability. There are some rings that are growing toward the nonlinear phase (see Figure \ref{fig:q2T20a1_sigd_st}). The radial drift of the rings halts as a result of further dust growth beyond $\taus=1$. On the other hand, remaining dust grains in gaps and other less dense regions still suffer from the radial drift toward the central star. For retaining such dust grains in the disk, either of the following two processes will be necessary: (A) coagulation instability of the remaining dust grains operates and triggers concentration into second generation rings, or (B) the first generation rings catch the remaining dust that radially drifts through the ring.

The process (A) will be expected since any disturbances on a disk give rise to seed perturbations; in the present simulations, we input seed perturbations only at $t=0\;\mathrm{yr}$. If dust grains in the first generation rings significantly grow in size ($\taus\gg 1$), coagulation instability of the remaining dust grains might be insensitive to the rings. Therefore, successive development of coagulation instability will play the key role in retaining dust grains in a disk.

Next we discuss the efficiency of the process (B). We evaluate collisional ``optical depth" regarding a collision between a drifting dust grain and dust grains in the first generation ring. The associated mean free path of a drifting dust at the midplane in a ring, $l$, is given by
\begin{equation}
l^{-1}=\frac{\Sigma_{\dst,\ring}}{\sqrt{2\pi}\hd(a_{\ring})m_{\mathrm{p},\ring}}\pi\left(a_{\ring}+a_{\drift}\right)^2,
\end{equation}
where $a_{\ring}$ and $a_{\drift}$ are the sizes of the dust in a ring and the drifting dust, respectively, and $m_{\mathrm{p},\ring}\equiv 4\pi\rhoint a_{\ring}^3/3$. We assume $\hd/H=\sqrt{\alpha/\taus}$ and the Epstein law, and thus $\tau_{\mathrm{s},\ring}/\tau_{\mathrm{s},\drift}=a_{\ring}/a_{\drift}$ in the ring. For a ring width of $\Delta R_{\ring}$, collisional ``optical depth", $\tau_{\mathrm{coll}}$, is $\sqrt{1+(\Delta v_{\phi}/\Delta v_r)^2}\Delta R_{\ring}/l$ and thus
\begin{align}
\tau_{\mathrm{coll}}\simeq&0.09\left(\frac{\Delta R_{\ring}/H}{0.02}\right)\left(\frac{\Sigma_{\dst,\ring}/\sigmag}{0.1}\right)\notag\\
&\times\left(\frac{\alpha}{10^{-4}}\right)^{-\frac{1}{2}}\left(\frac{\tau_{\mathrm{s},\ring}}{1}\right)^{-\frac{1}{2}}\left(1+\frac{\tau_{\mathrm{s},\drift}}{\tau_{\mathrm{s},\ring}}\right)^2\notag\\
&\times\sqrt{1+\frac{1}{4}\left(\frac{1 + \tau_{\mathrm{s,drift}}\tau_{\mathrm{s,ring}}^{-1} }{\tau_{\mathrm{s,ring}}^{-1}-\tau_{\mathrm{s,drift}}}\right)^2}.\label{eq:optical_depth}
\end{align}
The value of $\Delta R_{\ring}/H$ is roughly given by the value of the most collapsed ring in the q2T20a1 run: $\Delta R_{\ring}\sim 0.05\;\mathrm{au}$ and $H\simeq 2.17\;\mathrm{au}$ at $r\simeq32.3\;\mathrm{au}$. The value of $\tau_{\mathrm{s},\drift}/\tau_{\mathrm{s},\ring}$ in the last parenthesis on the second row of Equation (\ref{eq:optical_depth}) is less than unity, and thus we can neglect it. The last factor comes from $\sqrt{1+(\Delta v_{\phi}/\Delta v_r)^2}$ and is less than 10 for $\tau_{\mathrm{s,drift}}=0.1\ll\tau_{\mathrm{s,ring}}$. We then obtain $\tau_{\mathrm{coll}}<1$. The collisional optical depth can be much larger than unity only in a special case where the ring and the drifting dust have the similar radial velocity, i.e., $\tau_{\mathrm{s,drift}}\simeq\tau_{\mathrm{s,ring}}^{-1}$. This is not our main focus of this discussion since we consider dust grains that have radial relative velocity and drift with respect to the ring. In the limit of $\tau_{\mathrm{s,ring}}\gg1$, the last factor converges to $\sim 0.5/\tau_{\mathrm{s,drift}}\sim 5$ for $\tau_{\mathrm{s,drift}}=0.1$. In such a limit, we obtain $\tau_{\mathrm{coll}}\ll1$ because of the $\tau_{\mathrm{s,ring}}$-dependence appearing on the second row of Equation (\ref{eq:optical_depth}). In this way, Equation (\ref{eq:optical_depth}) indicates that the dust in the ring and the drifting dust are collisionless, i.e. the drifting dust grains can pass through the ring, especially after dust growth proceeds and $\tau_{\mathrm{s,ring}}$ becomes larger than unity. Therefore, we expect that the process (B) is inefficient in the present case.

As described above, the process (A) is more important to retain dust grains in a disk. Note that Equation (\ref{eq:optical_depth}) validates the hypothesis of the process (A) since it will occur when remaining dust grains are insensitive to large grains in the rings. For the further discussion, we need simulations with full-size distributions or at least two-population simulations that can treat drifting dust and large solids \citep[e.g.,][]{Drazkowska2017}. This will be addressed in our future studies.

\subsection{Comments on the development to secular GI}\label{subsec:commentSGI}
Our previous work proposed a scenario that, if coagulation instability operates in a disk, the instability is expected to set up preferable locations for secular GI to develop toward planetesimal formation. In this work, we demonstrate that coagulation instability operates indeed during the inside-out dust evolution. Those results support our scenario. As shown in Appendix \ref{app:sgi}, the rings observed in the present simulations can be unstable to secular GI if they form in outer region where Toomre's $Q$ is relatively small, for example $Q\lesssim 10$. Weaker turbulence is also required as shown in the previous studies on secular GI \citep[e.g.,][]{Youdin2011,Shariff2011,Takahashi2014,Tominaga2019}.

Although the present simulations demonstrate the possible path toward planetesimal formation, we have to treat the backreaction for rigorous discussions. The effect of the backreaction is expected to be insignificant in the linear growth phase. This is because coagulation instability starts to grow after dust density decreases and dust-to-gas surface density ratio becomes on the order of $10^{-3}$. On the other hand, the backreaction will become significant in the nonlinear growth phase, which will limit the maximum dust density in rings. We give further discussion on the development to secular GI in Paper II where we present simulations with not only the backreaction but also fragmentation.

\section{Summary}\label{sec:summary}
Formation of planetesimals has been widely studied but still under debate. In the case of outer planetesimal formation, disk instabilities have been proposed as a possible mechanism. In particular, some studies indicate that planetesimal formation via gravitational instability subsequent to streaming instability may explain the origin of Kuiper Belt Objects \citep[e.g.,][]{McKinnon2020}. However, the viability of such instabilities is still under debate because even weak turbulence ($\alpha\sim10^{-5}$) can suppress streaming instability \citep[e.g.,][]{McNally2021}. 

Our previous study \citep{Tominaga2021} proposed a new instability named coagulation instability, which is expected to trigger dust concentration and to assist planetesimal formation in a protoplanetary disk. Coagulation instability develops even in the presence of turbulence of $\alpha\sim10^{-4}$. 

In this work (Paper I), we perform numerical simulations and prove that coagulation instability operates during the inside-out evolution via dust growth and even for the background dust-to-gas ratio of $\sim10^{-3}$. The development of coagulation instability causes radial dust concentration into rings against the dust depletion due to the radial drift. Nonlinear growth also increases $\taus$ up to unity in the rings even at outer radii ($>10\;\mathrm{au}$), showing the possible path of dust growth beyond drift barrier. We intentionally exclude the backreaction and simplify the simulations to isolate the contribution to the dust concentration. The simplified simulations demonstrate that the backreaction is not prerequisite for the dust concentration via coagulation instability, which is in contrast to the self-induced dust trap \citep{Gonzalez2017}.

We measure the enhanced dust-to-gas ratio in the most collapsed ring at the time when the dimensionless stopping time $\taus$ becomes unity via coagulation instability. We find that the dust-to-gas ratio increases by an order of magnitude or more and that the ratio can reach $\sim0.1$ in the present simulations without backreaction. The resulting dust-to-gas ratio $\Sigma_{\dst,\ring}/\sigmag$ at the final time step is only slightly higher in a disk with larger $\eta\vk$ (Figure \ref{fig:etavk_dependence}). This is because background evolution leads to smaller $\taus$ (drift-limited), and coagulation instability increases dust density more until $\taus=1$ is achieved.

We also find that stronger turbulence slows the growth of coagulation instability, and as a result rings form at inner radii as $\alpha$ increases (Table \ref{tab:param} and Figure \ref{fig:cumumass_q2T20a1a5}). Too large $\alpha$ prevents coagulation instability from developing and creating dense rings before perturbations flow out of the numerical domain. 

We discuss the further development of resulting rings and dust grains in Section \ref{sec:discussion}. There are two possible ways to retain dust grains after the formation of the first generation rings: (A) coagulation instability of the remaining dust grains operates and form next generation rings, or (B) the first generation rings catch the remaining dust. We expect that the process (A) is more important for the dust retention. The process (B) is insignificant and dust grains coming from the outer region will pass though the ring without collisions once dust grains in the rings becomes large enough ($\tau_{\mathrm{s,ring}}\gg1$; see Equation (\ref{eq:optical_depth})).

\cite{Tominaga2021} proposed a scenario that a combination of coagulation instability and secular GI will explain planetesimal formation. The results of the present simulations support the scenario. As briefly mentioned in Section \ref{subsec:commentSGI}, the radial concentration observed in the present simulations will be the key to trigger secular GI in a disk (see also Appendix \ref{app:sgi}). Further discussion is given in Paper II.

In this paper, we neglect the effects of backreaction and dust fragmentation to isolate the contributions to dust growth and concentration. The simulations under this assumption are worth being presented since this simplification allows clear distinction between coagulation instability and the self-induced dust trap proposed in \cite{Gonzalez2017}: the backreaction is not prerequisite for the local dust concentration via coagulation instability. We expect that a combined process of coagulation instability and the self-induced dust trap is one promising mechanism to save dust grains (see discussion in Paper II). Linear growth phase is insignificantly affected by the backreaction since coagulation instability starts to grow after dust-to-gas ratio decreases ($\sim10^{-3}$). On the other hand, nonlinear growth of coagulation instability will be affected by the backreaction. The backreaction decelerates the radial drift and makes it easier for dust grains to overcome drift barrier in rings before $\sigmad/\sigmag$ increases significantly. Besides, the linear growth rate is reduced by fragmentation according to the linear analysis in \citet{Tominaga2021}. In Paper II, we present results of simulations where the effects of the backreaction (the drift deceleration) and the fragmentation are taken into account.

\acknowledgments
We thank Hidekazu Tanaka, Takeru K. Suzuki, Sanemichi Z. Takahashi and Elijah Mullens for fruitful discussions and helpful comments. We also thank the anonymous referee for the review, which helped us to improve the manuscript. This work was supported by JSPS KAKENHI Grant Nos. JP18J20360, 21K20385 (R.T.T.), 16H02160, 18H05436, 18H05437 (S.I.), 17H01103, 17K05632, 17H01105, 18H05438, 18H05436 and 20H04612 (H.K.). R.T.T. is also supported by RIKEN Special Postdoctoral Researchers Program.

%




\appendix
\section{Local simulations for weaker turbulence}\label{app:local_alp1e-5}
\begin{figure}[htp]
	\begin{center}
\raisebox{20pt}{
	\includegraphics[width=0.5\columnwidth]{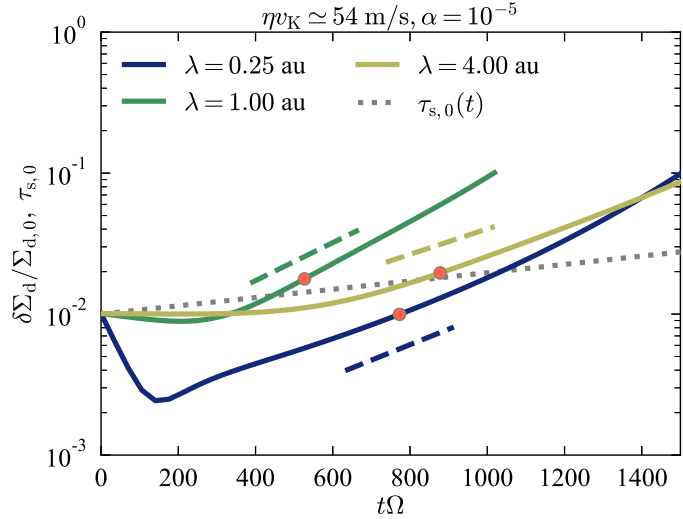}
	}
	\end{center}
	\vspace{-30pt}
\caption{Results of the local simulation with $\alpha=10^{-5}$ for $\lambda=0.25\;\mathrm{au},\;1\;\mathrm{au}$, and $4\;\mathrm{au}$. The blue, green and yellow solid lines show the evolution of $\delta\sigmad/\sigmadup$ for each wavelength. The gray dotted line shows time evolution of the background stopping time $\tausup$. The blue, green and yellow dashed lines show the growth rates derived in the previous linear analyses (Equation (\ref{eq:disp_app_diff})). We use physical values at time marked by the orange filled circles to derive the growth rate. The linear analyses have a good agreement with each simulation.}
\label{fig:local_1au_growth_lambdas_alpha1e-5}
\end{figure}

Figure \ref{fig:local_1au_growth_lambdas_alpha1e-5} shows results of the local simulations with $\alpha=10^{-5}$. We perform three runs with three different wavelengths: $\lambda=0.25\;\mathrm{au},\;1\;\mathrm{au},\;4\;\mathrm{au}$. As in the case of $\alpha=10^{-4}$, we find a good agreement between the previous linear analyses and the local simulations. The perturbations grows faster than in the simulations with $\alpha=10^{-4}$ because the diffusion is ineffective. The ratio of the growth rates and the coagulation rate (see the gray dotted line) is also larger. Thus, coagulation instability more quickly develops at shorter spatial scales in more weakly turbulent disks than assumed in Section \ref{sec:global}, leading to dust growth beyond the drift barrier. 

\section{Critical radius for coagulation instability to save dust grains}\label{app:Rc}
In the present simulations, we input initial perturbations only in the outer half region and observe that the perturbations can grow before they flow out of the numerical domain in most runs. This process helps dust grains to avoid radial drift and depletion. In this appendix, we roughly estimate a ``critical radius" that is a boundary dividing the radial regions where coagulation instability can save dust grains or not.

The necessary condition for the instability to save dust grains is that the radial drift timescale is longer than the growth timescale of coagulation instability. Similar discussion is given in Section 4.4 in \citet{Tominaga2021}. We here focus on the critical radius at which the ratio of the above two timescales is unity while we focused on $\taus$- and $\varepsilon$- dependences of the ratio in the previous study. Following the notation in \citet{Tominaga2021}, we denote the radial drift timescale by $t_{\mathrm{travel}}$, which is given by
\begin{equation}
t_{\mathrm{travel}}\equiv\frac{r}{|v_r|}.
\end{equation}
For simplicity, we assume $|v_r|\simeq 2\tausup\eta\vk$. The necessary condition to save dust is $t_{\mathrm{travel}}>n^{-1}$, which yields
\begin{equation}
r>\frac{2\tausup\eta\vk}{n}\simeq 11\;\mathrm{au}\left(\frac{\tausup}{0.1}\right)\left(\frac{\eta\vk}{54\;\mathrm{m/s}}\right)\left(\frac{n}{3\times10^{-3}\Omega(20\;\mathrm{au})}\right)^{-1}.\label{eq:critical_radii_save_dust}
\end{equation}
Note that the growth rate has $\tausup$- and $\eta\vk$-dependences. Roughly speaking, we can expect the growth rate to be proportional to the square root of those physical quantities. Thus, as a net dependence, the critical radius given the right hand side of Equation (\ref{eq:critical_radii_save_dust}) depends roughly on the square root of $\tausup$ and $\eta\vk$. We also note that the growth rate also depends on $\alpha$ and $\varepsilon$ \citep[see][]{Tominaga2021}.

Equation (\ref{eq:critical_radii_save_dust}) states that dust grains initially located at $r\simeq10\;\mathrm{au}$ are potentially saved by coagulation instability. The adopted value of $n$ in Equation (\ref{eq:critical_radii_save_dust}) is about the maximum growth rate for $\tausup=0.1,\;\alpha=1\times10^{-4}$ and $\varepsilon=1\times10^{-3}$ in the MMSN disk (e.g., see Figure 8 in \citet{Tominaga2021}). According to the results of the radial global simulations, coagulation instability already operates when $\varepsilon$ becomes just a few times lower than the initial value ($0.01$). In such cases, the growth rate is larger than the above value according to the linear analyses in \citet{Tominaga2021}, and the critical radius will decrease. On the other hand, the critical radius becomes larger if the radial diffusion is stronger since the growth rate decreases.

Note that perturbations that start growing at inner radii have larger growth rates since the growth rate is scaled by $\Omega\propto r^{-3/2}$, and thus the critical radius for such inner perturbations is smaller. For example, the critical radius is about 3.8 au for $n=3\times10^{-3}\Omega(10\;\mathrm{au})$, $\tausup=0.1$ and $\eta\vk=54\;\mathrm{m/s}$. In this case, inner dust grains will be saved as the inner perturbations grow via the instability.

Finally, we also note that Equation (\ref{eq:critical_radii_save_dust}) is the necessary condition. If the initial amplitudes of perturbations are small, it takes multiple growth timescales for coagulation instability to concentrates dust grains significantly. In such a case, the critical radius is larger than shown in Equation (\ref{eq:critical_radii_save_dust}).

\section{Secular Gravitational Instability in resulting dust dense regions}\label{app:sgi}

\begin{figure*}[tp]
	\begin{center}
	\raisebox{20pt}{
	\includegraphics[width=0.8\columnwidth]{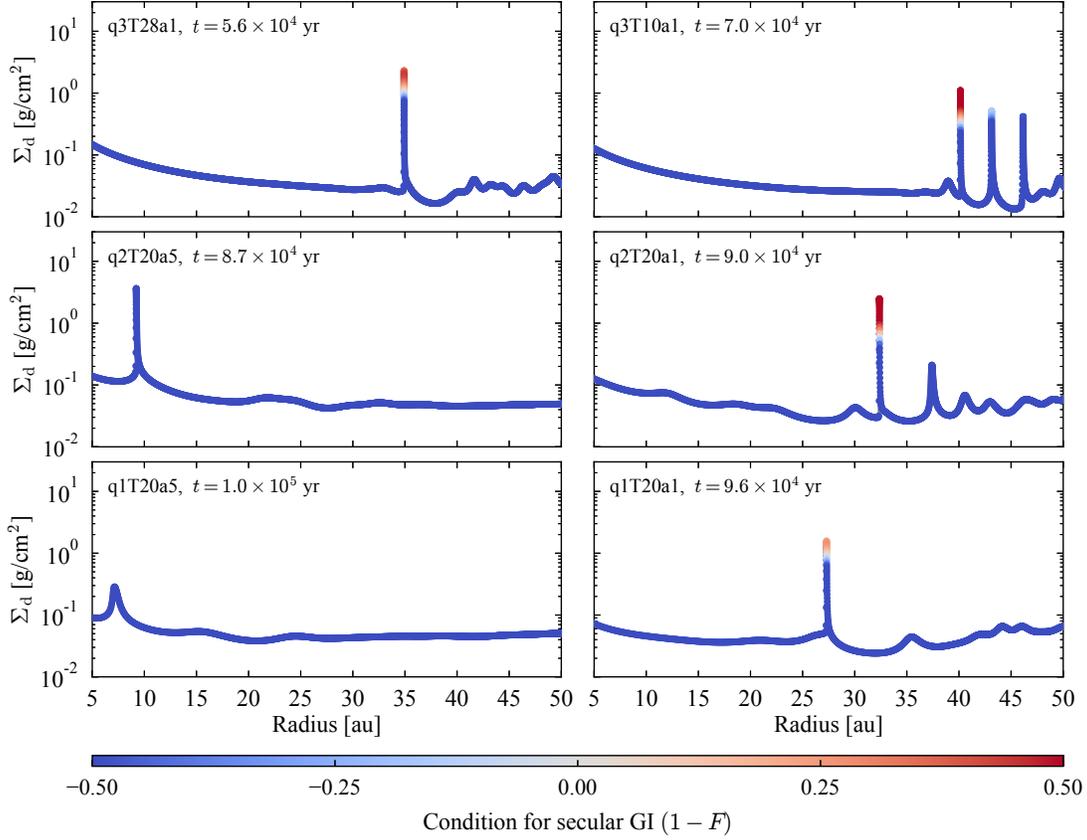}
	}
	\end{center}
	\vspace{-30pt}
\caption{Dust surface density profiles around the most collapsed ring in the six runs. We plot filled circles at the cell positions with color showing the condition for secular GI (see Equation (\ref{eq:SGIcond1})). Secular GI can operate for $1-F>0$. }
\label{fig:SGIcond_multiple}
\end{figure*}

The radial dust concentration shown in the present simulations is the possible connection to planetesimal formation via secular GI \citep[e.g.,][]{Ward2000,Youdin2005a,Youdin2005b,Youdin2011,Michikoshi2012,Takahashi2014,Takahashi2016,Tominaga2018,Tominaga2019,Tominaga2020,Pierens2021}. Although the backreaction should be taken into account in the simulations for more rigorous discussion (Section \ref{subsec:commentSGI}), it is still worth noting whether secular GI is operational in the rings observed in the present simulations without the backreaction. In this appendix, we thus briefly examine the stability of the rings.

Growth efficiency of secular GI is determined by four parameters: dust-to-gas ratio, dimensionless stopping time, turbulence strength, and Toomre's $Q$ value for gas. A condition for the growth of secular GI is given as follows \citep[][]{Tominaga2019}\footnote{We use the dust-to-gas surface density ratio to discuss the growth condition of secular GI although the dust-to-gas ratio around the dust sublayer might be more important. We do not discuss the effect of vertical structure in the context of secular GI because linear growth of secular GI on $r-z$ plane has not been studied and is beyond the scope of this paper.}:
\begin{equation}
F\equiv \frac{Q^2\left(\taus\cd^2/\cs^2+\tilde{D}\right)}{(1+\varepsilon)\left[\taus\left(\varepsilon+\cd^2/\cs^2\right)+\tilde{D}(1+\varepsilon)\right]}<1,\label{eq:SGIcond1}
\end{equation}
where $\tilde{D}\equiv D\cs^{-2}\Omega$ and $\cd\propto\sqrt{\alpha}\cs$ is velocity dispersion of dust grains \citep[see also][]{Youdin2005a,YL2007}. We briefly describe the physical insight into the growth condition of secular GI below. For $\tilde{D}\sim\cd^2/\cs^2\sim\alpha\ll\taus$ and $\tilde{D}\ll\taus\varepsilon$, Equation (\ref{eq:SGIcond1}) is reduced as follows (see also Equation (21) in \citet{Takahashi2014}):
\begin{equation}
F\simeq Q_{\mathrm{gd}}\times Q_{\mathrm{drag}}<1,
\end{equation}
\begin{equation}
Q_{\mathrm{gd}}\equiv\frac{\cs\Omega}{\pi G (\sigmag+\sigmad)},
\end{equation}
\begin{equation}
Q_{\mathrm{drag}}\equiv\frac{D/H^2}{\tstop\pi G\sigmad/H}=\frac{D\Omega}{\tstop\cs\pi G\sigmad}.
\end{equation}
This simplified equation clearly represents the key physical conditions for the onset of secular GI. The parameter $Q_{\mathrm{gd}}$ is a measure of how massive the dusty gas disk is. The other parameter $Q_{\mathrm{drag}}$ is a ratio of two timescales: (1) diffusion time over a spatial scale of $H$, and (2) a timescale for which dust moves over a distance $H$ at the terminal velocity $\tstop\pi G\sigmad$. Secular GI can grow in self-gravitationally stable disk with $Q_{\mathrm{gd}}>1$ when the dust diffusion is weak: $Q_{\mathrm{drag}}<1$ is necessary. Note that the diffusion coefficient $D$ in the linear analyses of secular GI is related to the radial diffusion. If the vertical diffusivity is also given by $D$, we obtain another form of the parameter $Q_{\mathrm{drag}}$ as follows
\begin{equation}
Q_{\mathrm{drag}}= \frac{\hd\Omega^2}{\pi G\sigmad}\times\frac{\hd}{H} =\frac{1}{\pi\sqrt{2\pi}} \frac{\Omega^2/ G}{\rho_{\dst}}\times\frac{\hd}{H},
\end{equation}
where $\rho_{\dst}=\sigmad/\sqrt{2\pi}\hd$ is the midplane dust density, and we also adopt $\hd=\sqrt{D/\taus\Omega}$ \citep[e.g.,][]{YL2007}. If we neglect the factor difference, the numerator of the first term, $\Omega^2/G$, roughly corresponds to the Roche-like density that determines the stability of three dimensional mode in a self-gravitating disk \citep[e.g., see][]{Goldreich1965a,Sekiya1983}. We can see that the above necessary condition $Q_{\mathrm{drag}}<1$ can be satisfied even for $\rho_{\dst}\lesssim\Omega^2/G$ since the dust layer is much thin ($\hd/H\ll1$) for grown dust grains. Therefore, secular GI does not require gravitationally unstable dust layer for its onset in contrast to the classical GI.

We include $\cd$ in the following discussion although our simulations does not include the effect of the velocity dispersion in the equation of motion for dust. Neglecting $\cd$ in Equation (\ref{eq:SGIcond1}) makes $1-F$ slightly larger, and a ring appears more unstable to secular GI. 

Figure \ref{fig:SGIcond_multiple} shows dust surface density profiles at the final time steps from the six runs. We overplot filled circles whose color shows whether the condition for secular GI is satisfied at the dust cells or not. Secular GI is operational at red-colored regions where $1-F$ is positive (Equation (\ref{eq:SGIcond1})). In some runs, we find that secular GI becomes unstable at the dust rings. 
 
The rings forming at outer radii can be sufficiently unstable to secular GI while inner rings tend to be stable or marginally unstable even when $\sigmad$ increases by an order of magnitude (e.g., see the panels of the q2T20a5 and q2T20a1 runs). This is partly due to difference in the turbulence strength $\alpha$. Dust diffusion with larger $\alpha$ stabilize secular GI. Thus, less turbulent disks are still preferable even when dust surface density is increased by coagulation instability. Toomre's $Q$ value of the background gas disk also significantly affect the stability of the rings. From the definition of Toomre's $Q$, we have
\begin{equation}
Q=\frac{\cs\Omega}{\pi G\sigmag}\propto r^{\frac{2q-7}{4}}\label{eq:tomq_r}
\end{equation}
for $T\propto r^{-1/2}$, $\Omega\propto r^{-3/2}$ and $\sigmag\propto r^{-q/2}$, meaning that $Q$ increases as $r$ decreases for $q=1,\;2$, and 3. As already mentioned in the previous studies, secular GI grows more easily for smaller $Q$ \citep[e.g., see][]{Youdin2005a,Youdin2011,Takahashi2014,Tominaga2019} although $Q<1$ is not required. In other words, secular GI becomes operational more easily at outer radii, and thus in outer rings. Note that secular GI is stable at dust-depleted regions (blue regions in Figure \ref{fig:SGIcond_multiple}). Thus, coagulation instability and resulting dust concentration are important for the onset of secular GI.

\section{On the effect of porosity of dust aggregates}\label{app:porosity}
Although we assume the spherical compact dust in the present simulations, dust collisions will produce porous aggregates \citep[e.g.,][]{Dominik1997,Suyama2008,Wada2008}. It is known that the coagulation timescale becomes shorter for such porous dust aggregation \citep[e.g.,][]{Okuzumi2012,Kobayashi2021}. The difference in the timescales can be estimated by introducing a filling factor $f\equiv \rho_{\mathrm{bulk}}/\rho_{\mathrm{mon}}$ where $\rho_{\mathrm{bulk}}$ and $\rho_{\mathrm{mon}}$ are the mass densities of a single aggregate and a single monomer, respectively \citep[e.g., see][]{Kataoka2013a}. As noted in the previous studies, the coagulation timescale is proportional to the mass-to-area ratio \citep[e.g., see Section 4 of][]{Okuzumi2012}:
\begin{equation}
\mpar\left(\frac{d\mpar}{dt}\right)^{-1}\propto \frac{\hd}{\Delta\vpp\sigmad} \frac{\mpar}{\pi a^2}\propto\frac{\hd}{\Delta\vpp\sigmad} f\rho_{\mathrm{mon}}a
\end{equation}
(see also Equation (\ref{eq:dmpdt})). Here, we assume that the mass of a single porous aggregate is given by $4\pi\rho_{\mathrm{bulk}}a^3/3$. \footnote{ As mentioned in \citet{Okuzumi2012}, the early stage of porous aggregation in a disk leads to aggregates whose fractal dimension is $\simeq2$. However, this stage will be irrelevant to coagulation instability. This is because the dimensionless stopping time of such aggregates is comparable to those of monomers and is too small, and they hardly drift. } Thus, the coagulation timescale is $f$ times shorter than in the case of a compact dust.

As found in \citet{Tominaga2021} and reviewed in Section \ref{subsec:review}, the growth timescale of coagulation instability is scaled by the coagulation timescale. Thus, coagulation instability operates even when we consider the porosity. In the porous-dust case, we expect that (1) the growth is faster and (2) the most unstable wavelength becomes shorter since the wavelength is scaled by the growth-drift length $L_{\gdl}=|v_r|\left(d\ln\mpar/dt\right)^{-1}$ \citep[][]{Tominaga2021}. We note that the stopping time is also proportional to $\mpar/\pi a^2$, and thus the growth-drift length is proportional to $f^2$.

According to the previous studies \citep[][]{Okuzumi2012,Kobayashi2021}, even porous aggregation is faced with the fast radial drift at $r>10\;\mathrm{au}$ while dust at $r\leq10\;\mathrm{au}$ quickly grows to planetesimals. Thus, such an outer region will be a place where coagulation instability significantly affect the dust evolution.

\bibliographystyle{aasjournal}
\bibliography{rttominaga2021b}

%
%
%


\end{document}